\documentclass{aa}
\usepackage{graphicx}
\usepackage{hyperref}
\usepackage{txfonts}
\usepackage{caption}
\usepackage{subcaption}
\usepackage{textcomp}
\usepackage{stfloats}
\usepackage{lipsum}
\usepackage{xcolor}
\usepackage{float}
\usepackage{natbib}
\usepackage[rightcaption]{sidecap}
\usepackage[displaymath,mathlines]{linenoaa}

\begin{document} 
 \makeatletter
 \let\linenumbers\relax
 \let\modulolinenumbers\relax
 \makeatother

\title{Three-part structure of solar coronal mass ejection observed in low coronal signatures of Solar Orbiter}
\titlerunning{Three-part structure of CME in Solar Orbiter}
\authorrunning{Podladchikova et al.}

\author{Tatiana Podladchikova \inst{1}
	\and Shantanu Jain \inst{1}
	\and Astrid M. Veronig \inst{2,3}   
	\and Stefan Purkhart \inst{2}
	\and Galina Chikunova \inst{5,1}
	\and Karin Dissauer \inst{4}
	\and Mateja Dumbovi\'c \inst{5}
}

\institute{Skolkovo Institute of Science and Technology, Bolshoy Boulevard 30, bld. 1, 121205, Moscow, Russia \\
	\email{t.podladchikova@skol.tech}
	\and
	University of Graz, Institute of Physics, Universit\"atsplatz 5, 8010 Graz, Austria
	\and
	University of Graz, Kanzelh\"ohe Observatory for Solar and Environmental Research, Kanzelh\"ohe 19, 9521 Treffen, Austria
	\and
	NorthWest Research Associates, 3380 Mitchell Lane, Boulder 80301, CO, USA
	\and
	Hvar Observatory, Faculty of Geodesy, University of Zagreb, Kaciceva 26, HR-10000, Zagreb, Croatia
}

	\date{Received Month, Year; accepted Month, year}

\abstract
{Coronal mass ejections (CMEs) are large-scale eruptions of plasma and magnetic field from the Sun propagating through the heliosphere. The March 28, 2022 event provides unique observations of a three-part solar coronal mass ejection (CME) in the low corona in active region AR 12975, which includes a bright core/filament, dark cavity, and bright front edge.}
{We investigate the relationship between coronal dimming, filament eruption, and early CME propagation in this rarely reported case. We employ 3D filament and CME shock reconstructions, alongside estimations of early CME evolution inferred from the associated expansion of the coronal dimming.} 
{We perform 3D reconstructions using data from Solar Orbiter (SolO), Solar TErrestrial RElations Observatory (STEREO-A), and Solar Dynamics Observatory (SDO) to analyse the path, height, and kinematics of the erupting filament. We develop and validate a method ATLAS-3D (Advanced Technique for single Line-of-sight Acquisition of Structures in 3D) against traditional approaches to reconstruct the filament loops and the CME shock structure. ATLAS-3D uses SolO data exclusively and integrates existing 3D filament reconstructions from the early stages of the event to establish spatial relationships between the filament and the CME frontal edge. Additionally, we employ the DIRECD method to estimate the characteristics of early CME propagation based on its coronal dimming evolution.}
{The filament height increased from 28 to 616~Mm (0.04 to 0.89~$\rm{R_{sun}}$) over 30 minutes from 11:05 to 11:35~UT, with a peak velocity of $648 \ \pm 51 \ \rm{km} \ \rm{s}^{-1}$ and a peak acceleration of  $1624 \ \pm 332 \ \rm{m} \ \rm{s}^{-2}$. At 11:45~UT, the filament deflected by 
about 12$^\circ$, reaching a height of 841~Mm (1.21~$\rm{R_{sun}}$). Simultaneously, the quasi-spherical CME shock expanded from 383 to 837~Mm (0.55 to 1.2~$\rm{R_{sun}}$) between 11:25 and 11:35~UT. Over 10 minutes, the distance between the filament apex and the CME leading edge more than doubled from approximately 93 to 212~Mm (0.13 to 0.3~$\rm{R_{sun}}$), demonstrating significant growth and increasing separation between them. Key parameters estimated from DIRECD and the 3D filament reconstructions include the CME direction (inclined by $6^\circ$ from radial expansion), a half-width of $21^\circ$, and a cone height of 1.12~$\rm{R_{sun}}$, which was derived at the end of the dimming's impulsive phase. The reconstructed 3D CME cone closely matches the observed filament shape at 11:45~UT in both height and angular width, representing its inner part. Validation with white-light coronagraph data confirmed the 3D cone's accuracy, particularly in matching filament and CME characteristics, including projections to STEREO-A COR2 times.}
{The study of the eruptive event on March 28, 2022, reveals rapid filament development and its subsequent deflection from the primary propagation direction. It confirms that connections between dimming and CME expansion can be established by the end of the dimming's impulsive phase, preceding the filament's deflection at 11:45~UT, illustrating further self-similar CME evolution. This approach links the expanding dimming with the early CME development, highlighting dimmings as indicators and suggesting the DIRECD method's utility in correlating the 2D dimming with 3D CME structure. These findings provide valuable insights into early CME evolution, and demonstrate the importance of multi-viewpoint observations and novel reconstruction methods for enhancing space weather forecasting. }


\maketitle

\section{Introduction} 
Coronal mass ejections (CMEs) are large-scale eruptions in the solar corona that involve the expulsion of vast quantities of magnetized plasma, with masses up to $10^{13}$~kg, driven by magnetic forces (e.g. reviews by \citealp{Forbes2006,Vrsnak2008,Chen2011,Webb2012,Cheng2017,Green2018}). CMEs are expelled from the Sun into interplanetary space with speeds ranging from approximately 100 to 3500~km/s \citep{Gopalswamy2009,Michalek2009,Tsurutani2014,Veronig2018,RodriguezGome2020}. They are the most energetic events in our solar system, capable of releasing energy up to approximately $10^{32}$ erg \citep{Vourlidas2010,Emslie2012}.

When the interplanetary counterpart of a CME, the ICME, and its associated shock reach Earth, they can cause strong geomagnetic storms \citep{Gonzalez1994,Tsurutani1992,Podladchikova2012,Podladchikova2018} and numerous space weather effects in geospace, such as power grid failures, radio signal disruptions, and various satellite malfunctions, making it crucial to detect and study their early evolution \citep{Sandford1999,Doherty2004,Pulkkinen2007,Baker2013,Koskinen2019,Temmer2021}. However, coronagraphs often do not catch the early stages of the CME evolution due to occultation or projection effects \citep{Gopalswamy2000,Burkepile2004,Schwenn2005}. Analysing CME-related phenomena in the low corona, such as filament eruptions and coronal dimmings, is thus essential for understanding the initiation mechanisms and early evolution of CMEs. Moreover, recent studies have shown that widespread solar energetic particles (SEPs) can fill the entire heliosphere, which may be related to the properties of the CME and its associated shock in the low corona \citep{Dresing2012,Gomez-Herrero2015,Lario2016,Dresing2023}.

CMEs observed in white-light coronagraphs often display a three-part structure: a leading bright front, a dark cavity, and a bright core \citep{Illing1986}. While these features are well-known, cases where the CME three-part structure is clearly seen in extreme ultraviolet (EUV) \citep{Balmaceda2022} are rarely reported. The bright front is a shell of compressed material, while the dark cavity and bright core indicate a magnetic flux rope, with the core often thought to be cool, dense prominence material \citep{Low1995,Dere1999,Labrosse2010,Parenti2014,Gibson2015,Gibson2018,Chen2020}. However, \citet{Howard2017} suggested the core represents the erupting flux rope itself. \citet{Veronig2018} reported for the X9 flare on 2017 September 8 that the hot, bright rim around the expanding EUV cavity, formed by poloidal flux and heated plasma added through magnetic reconnection, extends into the coronagraph field-of-view (FOV) as the CME core, supporting the view that in this case the core represents the flux rope formed by the eruption, and not trapped prominence material. 

In recent years, a number of studies have focused on the CME trajectory determination and 3D CME reconstructions using dual and triple coronagraph observations from the STEREO and SOHO spacecraft. Techniques such as tie-pointing \citet{Hildner1977,MacQueen1986,Byrne2010,Liu2010,Liewer2011,Isavnin2013,Liewer2015}, epipolar geometry \citep{Inhester2006,Podladchikova2019}, and filament reconstruction \citep{Illing1986,Liewer2009,Li2013} have been used to determine CME and filament positions in 3D. Forward modelling, including the Graduated Cylindrical Shell (GCS) model \citep{Thernisien2006,Thernisien2011}, has also been applied, linking solar observations with in situ signatures at 1~AU \citep[e.g.][]{Gui2011,Kay2016,Temmer2017,Kay2024}. 

Coronal dimmings, regions of strongly reduced EUV and X-ray emissions due to CME expansion and resulting density depletion, are closely linked to CME evolution \citep{Hudson1996,Sterling1997,Thompson1998}. This relationship is supported by Differential Emission Measure studies revealing distinct density drops \citep{Cheng2012,Vanninathan2018,Veronig2019} and spectroscopic observations of strong outflows \citep{Harra2001,Harra2003,Jin2009,Tian2012} in dimming regions. Several studies found distinct correlations between CME mass and dimming area, brightness, and magnetic flux, as well as between CME speed and the time derivatives of these parameters \citep{Harrison2000,Harrison2003,Zhukov2004,Lopez2019,Wang2023,Baker2013}. Other research has examined the morphology and early evolution of the CME  \citep{attrill2006using, qiu2017gradual}, along with their timing \citep{miklenic2011coronal, ronca2024recoverycoronaldimmings}. Additionally, statistical analyses have underscored the importance of coronal dimmings in understanding CME dynamics \citep{Bewsher2008,Reinard2009,Mason2016,Krista2017,Dissauer2018a,Dissauer2018b,Dissauer2019,Chikunova2020}. A number of studies highlight the importance of coronal dimmings for studying CME propagation and deflection. \citet{Thompson2000} were the first to point out that the extended dimming areas map to the CME footprints. \citet{Chikunova2023} linked the dimming morphology to 3D CME structure, suggesting that the dimming observations can provide insight into the CME direction. Recently, \citet{Jain2024,jain2024_may_june_storm} developed a method, called DIRECD, to estimate the CME direction from the expansion of the coronal dimmings.

In this study, we present a rarely reported case of a three-part CME observed in the low corona on March 28, 2022, characterized by a bright core, dark cavity, and bright front edge. The event is a textbook eruptive two-ribbon flare with filament eruption and associated high-energy HXR emission \citep{Purkhart2023}. Interestingly, the filament has formed only about one day before, and underwent a fundamental topological restructuring (changing one of its footpoints), which led to its eruption and the associated M4 flare about 1.5 hours afterwards \cite{Purkhart2024}. In this paper, we investigate the global 3D aspects of the eruption, making full use of the simultaneous multi-point observations from three spacecraft with different viewpoints. We perform 3D filament and CME shock reconstructions from three vantage points provided by the SolO, STEREO-A, and SDO spacecraft, and also employ the DIRECD method to infer early CME evolution from the associated coronal dimming.

\section{Three-dimensional properties of the erupting filament and the CME shock}
This section details the observation of the March 28, 2022 filament eruption and CME shock structure, revealing the rarely observed three-part structure of the CME in the low corona: a bright core/filament, a dark cavity, and a bright CME front edge. We first perform 3D reconstructions to analyse the filament's path, height, and kinematics, highlighting the rapid development and propagation of the eruption using the three different vantage points by SolO, STEREO-A, and SDO (Section~\ref{3d_filament_tips}). We further reveal the deflection of the filament from its primary propagation direction and introduce a new reconstruction method called ATLAS-3D (Advanced Technique for single Line-of-sight Acquisition of Structures in 3D) to derive the 3D structure of filament loops and the CME shock using data solely from the SolO spacecraft but incorporating the existing 3D filament reconstructions from Section~\ref{3d_filament_tips}. This approach allows us to establish spatial relationships between the filament and the CME's frontal edge. Additionally, we validate the method against traditional 3D approaches, reconstructing the observed CME shock in various wavelengths by SolO and STEREO-A (Section~\ref{3d_shock}).

\subsection{Three-dimensional reconstructions of the filament height evolution}\label{3d_filament_tips}
\begin{figure}  
	\centering
	\includegraphics[width=1\columnwidth]{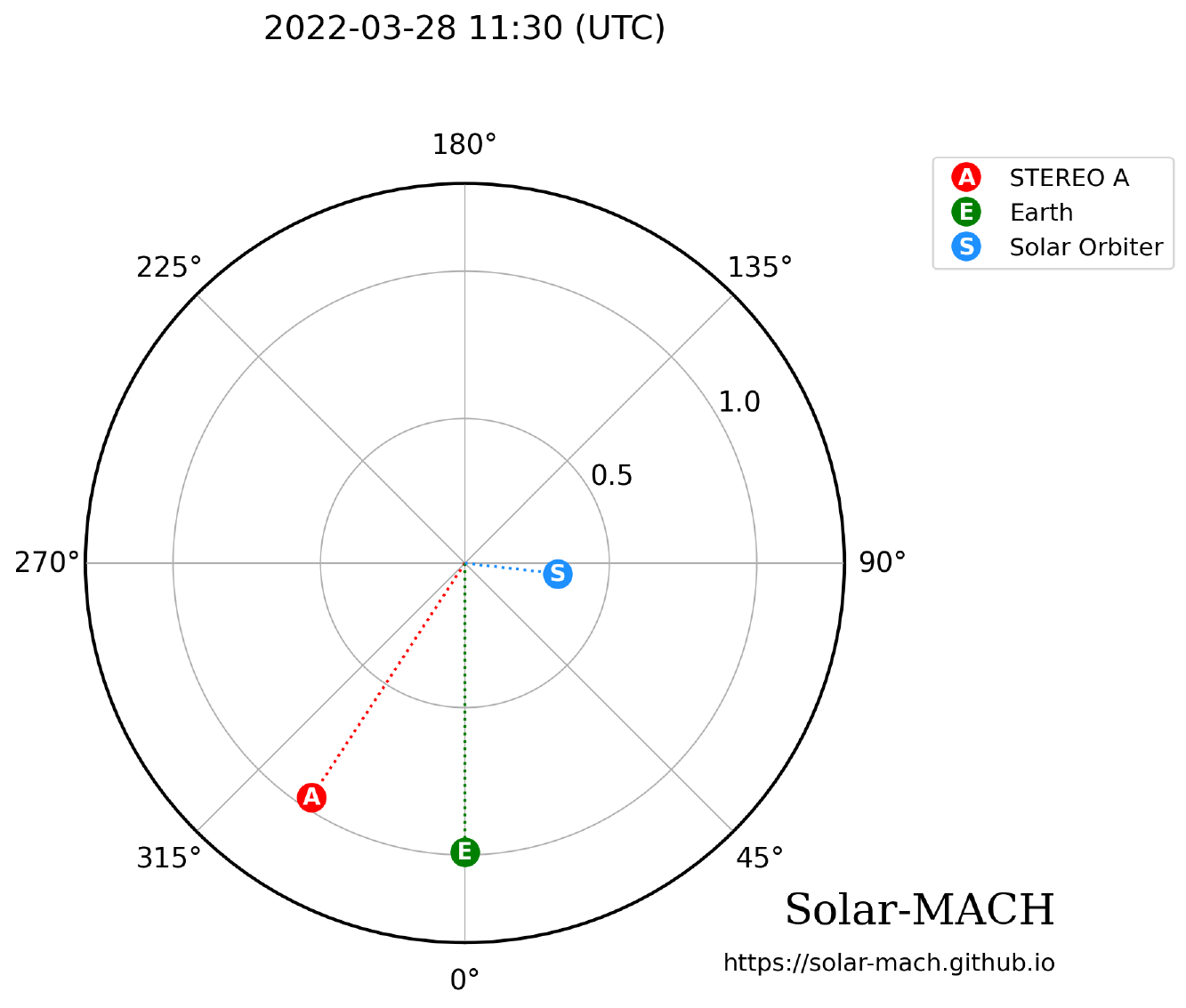} 
	\caption{Location of the three observing 
        spacecraft on 28 March 2022. STEREO-A and Earth (SDO) are separated by $33^\circ$ in longitude, while SolO and Earth (SDO) are separated by $83.5^\circ$. Visualisation done by the Solar MACH tool \citep{Gieseler2023}.}
	\label{fig1}
\end{figure}
\begin{figure*}
 \centering
 \includegraphics[width=16cm]{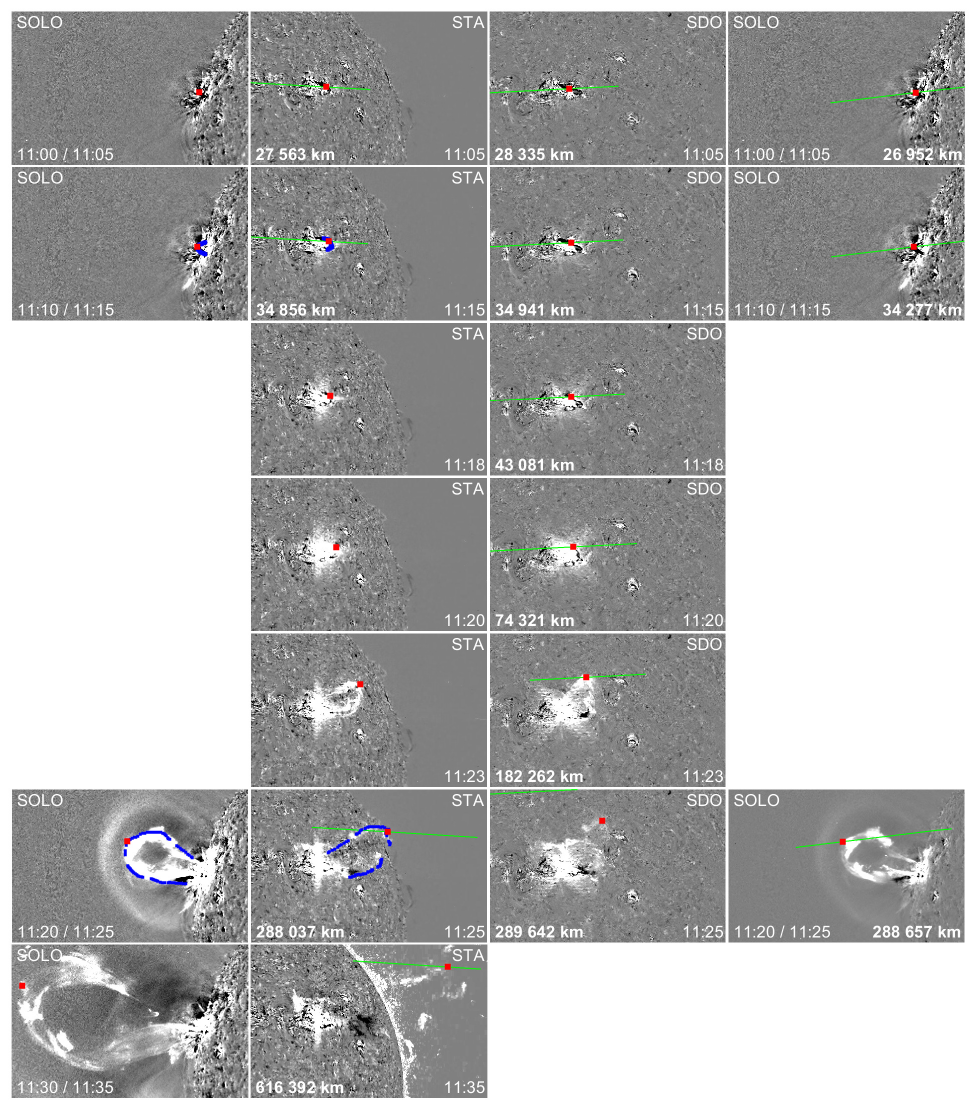}
 \caption{3D reconstruction of the erupting filament between 11:05 and 11:35~UT in SolO/EUI (SOLO), STEREO-A/EUVI (STA) and SDO/AIA 304~{\AA} base difference images. Column 1 includes both SolO and Earth time for the convenience to compare with Earth-based observers. Red marker shows the upper tip of the filament matched at three viewpoints. Epipolar lines are shown in green.  Blue markers indicate the selected points along the filament used for the 3D reconstructions of the filament loops at 11:15 and 11:25~UT in SOLO and STEREO-A base difference images.} \label{fig2}
\end{figure*}
\begin{figure}  
	\centering
	\includegraphics[width=0.8\columnwidth]{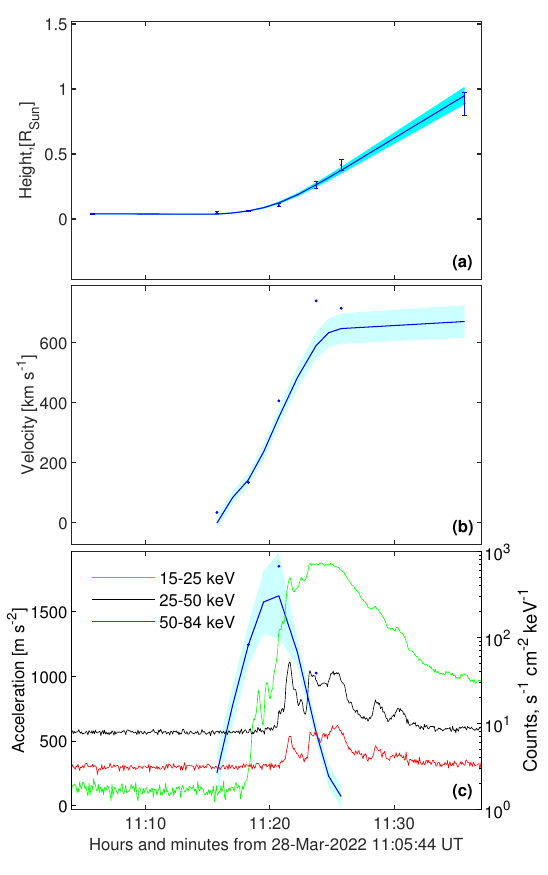} 
	\caption{Kinematics of the erupting filament and the SolO/STIX HXR flux of the associated flare in three energy bands from 15 to 84~keV.	
		(a) Filament height-time plot (blue circles) obtained from SolO, STEREO-A and SDO 304~\AA~images together with error bars (10\% from the height values). 
		The corresponding solid line shows the smoothed height–time profile. (b) Velocity and (c) acceleration of the filament obtained by numerical differentiation of the measured height-time data (dots) and the smoothed profiles (lines). The shaded areas outline the error ranges. SolO/STIX data are shifted by the difference in travel time ($+319$ sec) with respect to Earth.}
	\label{fig3}
\end{figure}
The filament eruption on March 28, 2022, is particularly interesting due to its observation from three separate vantage points by the SolO, STEREO-A, and SDO spacecraft. Figure~\ref{fig1} schematically illustrates the locations of the three spacecraft \citep{Gieseler2023}, with a longitudinal separation of $33^\circ$ east for STEREO-A and $83.5^\circ$ west for SolO from the Sun-Earth line. The M4 flare and filament eruption took place in Active Region 12975, located close to the central meridian as seen from the Sun-Earth line. SolO was near its first perihelion during the nominal science phase, positioned only 0.33~AU from the Sun. As a result, it takes approximately 162 seconds for a light signal to reach SolO, compared to around 482 seconds for Earth-based observers (0.966~AU for STEREO-A, 0.998~AU for SDO). Thus, to synchronize the filament dynamics observed by all three spacecraft, we must account for a time shift of approximately 319 seconds in STEREO-A and SDO images relative to the SolO observing time. This adjustment is particularly crucial when the filament eruption develops rapidly. 

By taking advantage of the available multi viewpoint observations, we derive the 3D reconstructions of the propagating filament path and heights using the epipolar geometry approach (see a detailed description in \citealp[][]{Thompson2009, Inhester2006, Podladchikova2019}). Figure~\ref{fig2} shows the evolution of the erupting filament between 11:05 and 11:35~UT in 304~\AA~SolO/EUI (columns~1~and~4), STEREO-A/EUVI (column~2) and SDO/AIA (column~3) base difference images. All data series were calibrated and corrected for differential rotation to a reference time of 28 March 2022, 10:30~UT using Solarsoft. In addition, we enhanced the counts of off-disk pixels in SolO/EUI images using a radial filter based on a simple sigmoid function. This was done to improve the simultaneous visualization of on- and off-disk emission.

From the SolO perspective, the event was located near the eastern limb of the Sun, whereas STEREO captured it in the north-western part of the Sun closer to the limb, and SDO near the disk centre. To facilitate comparison of SolO and Earth-based observations, we include both SolO and Earth time in column~1. We further utilize Earth-based observer time elsewhere in the paper. At each time step, we first manually identify the highest points of the filament (red markers) in SolO images (Figure~\ref{fig2}; column 1)
and match these with the corresponding points on the epipolar lines (green) in STEREO-A images (column~2). We further match the upper tips of the filament observed in STEREO-A (column~2) with the highest filament points seen in SDO (column~3). These points are then matched with those captured by SolO (column~4). Utilizing these three viewpoints provides an additional check on the accuracy of the 3D reconstructions. For each stereoscopic pair of spacecraft, we obtain 3D coordinates for the highest points of the filament, defining its position in 3D. We then determine the filament heights for each time step, showing the filament growth from approximately 28 to 616~Mm (0.04 to 0.89~$\rm{R_{sun}}$) over 30 minutes, between 11:05 and 11:35~UT, while considering a relative error of 10\% in the height values, estimated by the ten-pixel shift used to find the matching point along the epipolar line. Additionally, by constructing a linear fit through the filament's highest points, we  
obtain a nearly radial filament motion with an inclination of $6^\circ$ from the radial direction. We further mapped the filament direction onto the meridional and equatorial planes, calculating the 2D angles between the projections in each plane. The red meridional plane contains the Sun's centre O, the event source C on the solar surface, and the North Pole. The green plane passes through the source region and is parallel to the solar equator. The analysis shows that the filament deviates from the radial direction by $5.6^{\circ}$ to the south in the meridional plane and by $2.8^{\circ}$ to the west in the equatorial plane.

The filament heights obtained from the 3D reconstructions of the filament's upper tip are then used to study its kinematics. Figure~\ref{fig3} illustrates the time evolution of the filament height (a), velocity (b), and acceleration (c), compared to the STIX hard X-ray count rates at energies of 15--25 (green), 25--50 (black), and 50--84 (red) keV. In panel a, the filament heights are plotted as a function of time (blue circles), along with error bars. To derive the filament's velocity and acceleration profiles, we first smooth the filament height data and then compute the first and second numerical derivatives. The smoothing technique, as presented in \citet{Podladchikova2017} and extended to non-equidistant data, optimises the balance between data fidelity and curve smoothness (see applications in \citealp[][]{Veronig2018,Dissauer2019,Gou2020,Saqri2023,Chikunova2023}). Using the derived acceleration profile, we interpolate to equidistant data points (solid line in panel c), minimising the second derivatives to achieve a smoother resulting curve, and reconstruct the corresponding velocity (solid line in panel b) and height (solid line in panel a) profiles by integration. We also estimate the errors in the kinematic profiles by representing the reconstructed filament height, velocity, and acceleration as explicit functions of the original errors in the height-time data (blue shaded areas). The dots in panels b and c represent the first and second time derivatives obtained directly from the height measurements. As shown in the kinematic profiles in Figure~\ref{fig3}, the filament exhibits a very rapid evolution. The velocity continuously increases, reaching a peak of  $648 \ \pm 51 \ \rm{km} \ \rm{s}^{-1}$ at a distance of  222~Mm (0.32~$\rm{R_{sun}}$) above the limb at 11:24:43~UT. The acceleration profile reaches its maxima at 11:19:29~UT with a value of $1624 \ \pm 332 \ \rm{m} \ \rm{s}^{-2}$. As observed in panel c, the peak of the filament acceleration precedes the first flare HXR peak by about 2--3 min. The overall phase of increasing filament velocity (panel b) roughly aligns with the first phase of the impulsive HXR flare emission. SolO/STIX data are also shifted by the difference in travel time ($+319$ sec) with respect to Earth.

\subsection{Three-dimensional reconstructions of filament loops and CME shock}\label{3d_shock}
\begin{figure*}
   \centering
    \begin{subfigure}[b]{\textwidth} 
		\centering
        \includegraphics[width=0.84\textwidth]{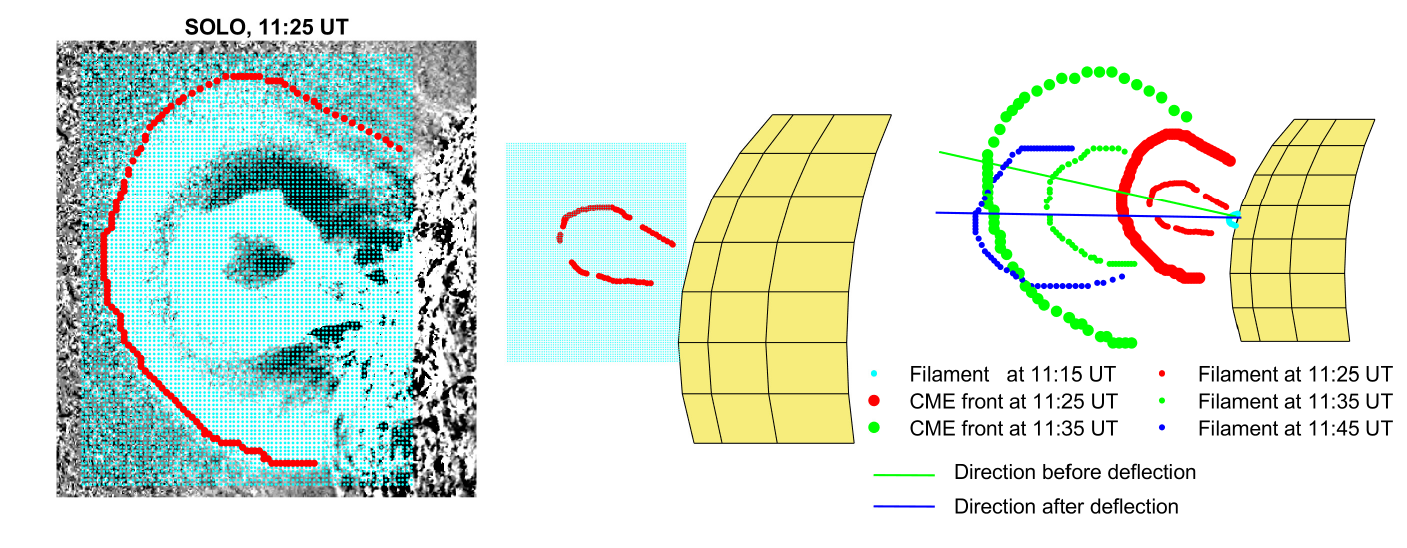}
		\caption*{}
		\label{}
	\end{subfigure}    

    \begin{subfigure}[b]{\textwidth} 
		\centering
        \includegraphics[width=0.84\textwidth]{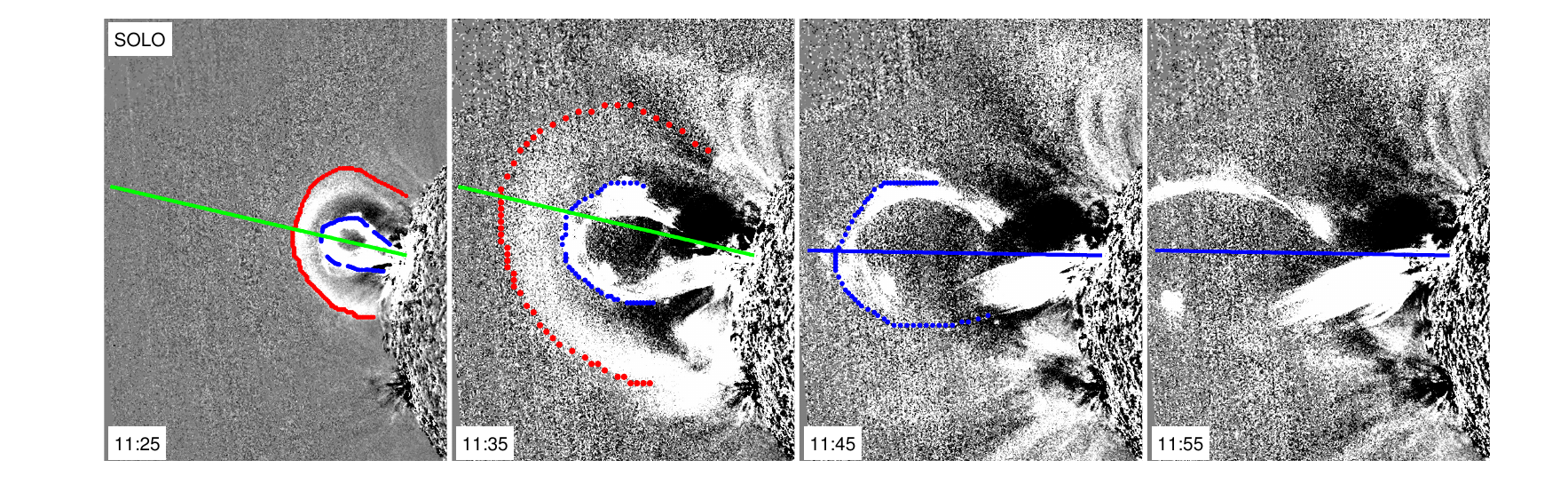}
		\caption*{}
		\label{}  
	\end{subfigure}

    \begin{subfigure}[b]{\textwidth} 
		\centering
		\includegraphics[width=0.84\textwidth]{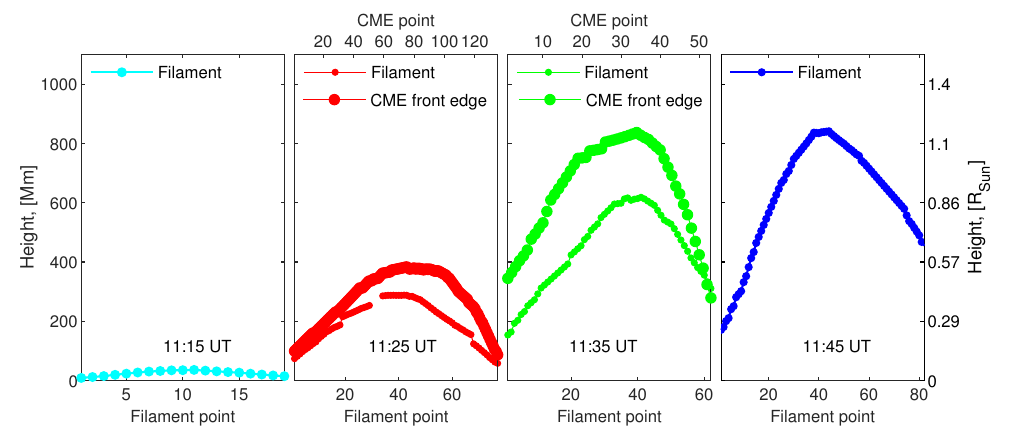}
        \caption*{}
		\label{}  
	\end{subfigure}
	\caption{Full 3D reconstructions of the outer CME front and filament structures. Top row: ATLAS-3D reconstructions of the outer CME front  and the filament. The left panel shows LOS projections of a 3D plane (cyan mesh) constructed through the 3D filament points (red markers) in the 304~\AA~SolO/EUI base-difference image. The 3D plane itself (cyan mesh), along with the 3D filament coordinates (red markers), is shown in the middle panel. The right panel gives the 3D coordinates of all considered filament and CME front structures, employing both 3D reconstructions based on epipolar geometry and the ATLAS-3D method. At 11:45~UT, a filament deflection on $11.9^\circ$ is observed. The green line indicates the filament pre-deflection direction, determined as the linear fit through its highest points, while the blue line represents its post-deflection direction. Middle row: Evolution of the filament as seen in SolO/EUI 304~{\AA} base-difference images over 30 minutes from 11:25 to 11:55~UT. Blue dots outline the filament, red dots the CME front. The green line depicts the primary direction of filament propagation, with a deflection (indicated by the blue line) observed at 11:45~UT. Bottom row: Heights of the filament loops and outer CME front edges from 11:15 to 11:45~UT. The bottom x-axis represents a filament point, while the top x-axis (for 11:25 and 11:35 UT) shows a point of the CME's front edge.}
	\label{fig4}
\end{figure*}
\begin{figure*}  
	\centering
    \includegraphics[width=1\textwidth]{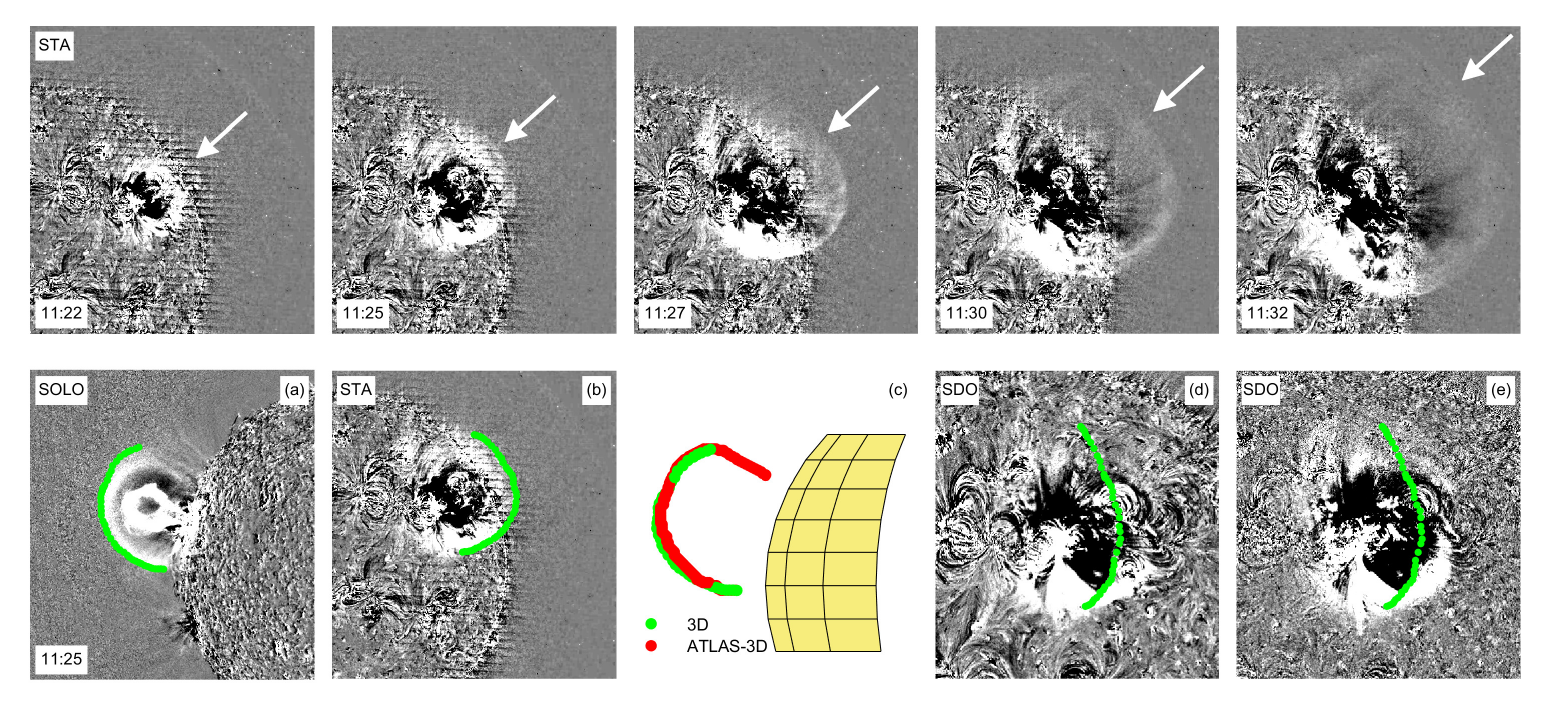}	
	\caption{Quasi-spherical CME shock and its 3D Reconstruction in multi-instrument views. Top row: Quasi-spherical CME shock (indicated with arrows) as seen from base-difference STEREO-A 195~\AA~from 11:22 to 11:32~UT. Bottom row: 3D reconstructions of the outer CME front based on epipolar geometry and two viewpoints from SolO and STEREO-A. (a) SolO/EUI 304~{\AA} and (b) STEREO-A/EUVI 195~{\AA} views in base-difference images. (c) 3D coordinates of reconstructed CME front outer edge based on epipolar geometry (green markers) and ATLAS-3D method (red markers). (d) LOS projections of 3D coordinates of the CME front outer edge to the SDO/AIA {195~\AA} view in base-difference and (e) running-difference images. }
	\label{fig5}
\end{figure*}
SolO and STEREO-A multi-point views in 304~\AA~allow us not only to estimate the filament's heights but also to perform 3D reconstructions of the specific filament loops observed at 11:15 and 11:25~UT on both spacecraft, shown in Figure~\ref{fig2}. The blue markers outline selected points along the filament used for the 3D reconstructions. The SolO/EUI image at 11:25~UT reveals both the filament loop and another, leading front, a phenomenon rarely observed in EUV 304 {\AA} imagery. This image clearly demonstrates a rarely reported case of the CME three-part structure observed in the low corona, consisting of a bright core/prominence, a dark cavity, and a bright CME frontal edge. While it is clear in this image that the prominence/filament corresponds to the bright core, it is not clear whether the leading front corresponds to the CME compressed material or a local density enhancement due to the shock. We do note however, that the quasi spherical dome-shape of the front (and its correspondence in the STEREO-A/EUVI 193 {\AA} images, to be discussed further) supports the shock scenario better. Interestingly, while this front is visible in the 304~\AA~ SolO/EUI, it is not seen in the 304~{\AA} STEREO-A/EUVI images. This discrepancy could be due to the SolO's closer proximity to the Sun, resulting in a better photon statistics and higher sensitivity.

To resolve the origin and evolution of the observed EUV~304~{\AA}~front, we propose a new method (ATLAS-3D)  
that leverages existing 3D filament reconstructions, as is illustrated in the first row of Figure~\ref{fig4}. First, we construct a 3D plane (cyan mesh, middle panel) through the 3D filament points (red markers). Second, we project the resulting 3D plane onto the SolO/EUI 304~{\AA}~base-difference image (left panel) along the line of sight (LOS). Finally, we manually select the 3D coordinates that correspond to the outer edges of the CME front. We repeat the same procedure for later time steps, reconstructing both the CME front and filament loop at 11:35~UT, as well as the filament loop at 11:45~UT. The 3D coordinates of all the considered structures, employing both 3D reconstructions based on epipolar geometry using multiple viewpoints and the ATLAS-3D method, is displayed in the right panel. LOS projections of the resulting 3D coordinates of CME fronts (red markers) and filament loops (blue markers) are provided in the second row of Figure~\ref{fig4}. As seen from both the first and second rows of Figure~\ref{fig4}, at 11:45~UT the filament gets deflected (blue lines) from the primary direction of its propagation (green lines) by $11.9^\circ$, suggesting that after the initial deviation, there is no further deflection and the CME expands self-similarly.

The third row of Figure~\ref{fig4} shows the heights of all examined filament loop and CME front structures. At 11:15~UT, the filament reached an apex height of approximately 36~Mm (0.05~$\rm{R_{sun}}$), swiftly increasing to 288~Mm (0.41~$\rm{R_{sun}}$) around 11:25~UT, with the front edge appearing at 383~Mm (0.55~$\rm{R_{sun}}$) at the same time, possibly as the bow shock was formed during this impulsive phase of the eruption. By 11:35~UT, the filament had grown to about 616~Mm (0.89~$\rm{R_{sun}}$), and the front outer edge to 837~Mm (1.2~$\rm{R_{sun}}$). Further filament growth to 841~Mm (1.21~$\rm{R_{sun}}$) was observed at 11:45~UT. We note that the distance between the filament and the CME front outer edge along the filament propagation direction (Figure~\ref{fig4}, first row, right panel, green line) more than doubled from around 93 to 212~Mm (0.13 to 0.3~$\rm{R_{sun}}$) over 10 minutes, from 11:25 to 11:35~UT, showing significant growth and increasing separation between them. 
\renewcommand{\arraystretch}{1.2} 
\begin{table}[h]
    \centering
    \caption{Filament loop and outer CME front heights}
    \begin{tabular}{l c c c c} 
        \hline
        \textbf{Time} & \multicolumn{2}{c}{\textbf{Filament height}} & \multicolumn{2}{c}{\textbf{CME height}} \\  
         & Mm & $\rm{R_{sun}}$ & Mm & $\rm{R_{sun}}$ \\  
        \hline 
        11:15 & 36 & 0.05  &     &      \\
        \hline 
        11:25 & 288 & 0.41 & 383 & 0.55  \\      
        \hline 
        11:35 & 616 & 0.89 & 837 & 1.2   \\      
        \hline 
        11:45 & 841 & 1.21 &     &    \\      
        \hline 
    \end{tabular}
    \label{table1} 
    \tablefoot{Before 11:45~UT, the filament propagated nearly radially with a 6$^\circ$ tilt, deflecting by 12$^\circ$ at that time.}
\end{table}

To validate the suggested ATLAS-3D reconstruction method, we applied it to estimate the filament's highest point at 11:35~UT, resulting in a value of 618~Mm, which closely matches the height of 616~Mm (0.89~$\rm{R_{sun}}$) obtained from 3D reconstructions based on epipolar geometry using two viewpoints (SolO and STEREO-A). Additionally, the 3D distance between the coordinates of the filament's tip (its apex) obtained from the two methods is only 6.6~Mm (0.009~$\rm{R_{sun}}$) with a relative error of 0.3\%, affirming the validity of the suggested ATLAS-3D method, especially given the filament's motion close to radial and its further deflection in the same plane.

Finally, we can also observe what appears to be a quasi-spherical CME shock from an additional vantage point, as demonstrated in the first row of Figure~\ref{fig5}. It shows STEREO-A/EUVI 195~{\AA} base-difference images from 11:22 to 11:32~UT, where we see enhanced propagating emission fronts in the low corona \cite[a  typical signature of large-scale EUV waves;][]{Warmuth2015}, in particular toward the South, but part of the emission is also observed off-limb in a 3D dome-like shape that directly connects to the low-lying EUV wave fronts. Note that such an observation is a rare case of a fully dome-shaped EUV wave indicative of the coronal shock associated with the eruption \citep{veronig2010,kozarev2011}. We also highlight the sharp front that is particularly well seen at the frame at 11:27~UT. 
Considering the sharp edges observed in the quasi-spherical dome, which are characteristic of a discontinuity rather than CME loops \citep{veronig2010}, along with the detection of a Type II solar radio burst, a signature of a coronal shock \citep{Ndacyayisenga2024,Morosan2024}, 
we suggest that the leading front observed in SolO is indeed a shock. The type II burst starts at 11:23 UT, which roughly coincides with the appearance of the EUV dome, and the radio imagery from the Nancay Radio Heliograph in \citep[][see their Fig.~3]{Morosan2024} indicates that the type II source location is above/at the EUV dome.

To further support this interpretation, we match the outer edge of the CME front in the SolO/EUI 304~{\AA} image to that in the STEREO-A/EUVI 195~{\AA} image using epipolar geometry. The results are shown in the second row of Figure~\ref{fig5}, where green markers in panel a (SolO) and panel b (STEREO-A) indicate the leading front's outer edge, matching the same features on both spacecraft. Panel c depicts closely matching 3D coordinates of the reconstructed leading front outer edge based on epipolar geometry (green markers) and the ATLAS-3D method (red markers), supporting both the observation of the CME shock and the effectiveness of the suggested ATLAS-3D method. We also present LOS projections of the estimated 3D coordinates of the CME front outer edge to the SDO view in AIA 195~{\AA} base-difference (panel d) and running-difference images (panel e, where the signal is more prominent), showing that 3D reconstructions with SDO would be challenging in this case due to the LOS integration of the optically thin emission against the solar disk. 

\section{Relationship between the coronal dimming, filament eruption and early CME propagation}
\begin{figure}
   \centering
    \begin{subfigure}[b]{0.46\textwidth}
       \centering
        \includegraphics[width=\textwidth]{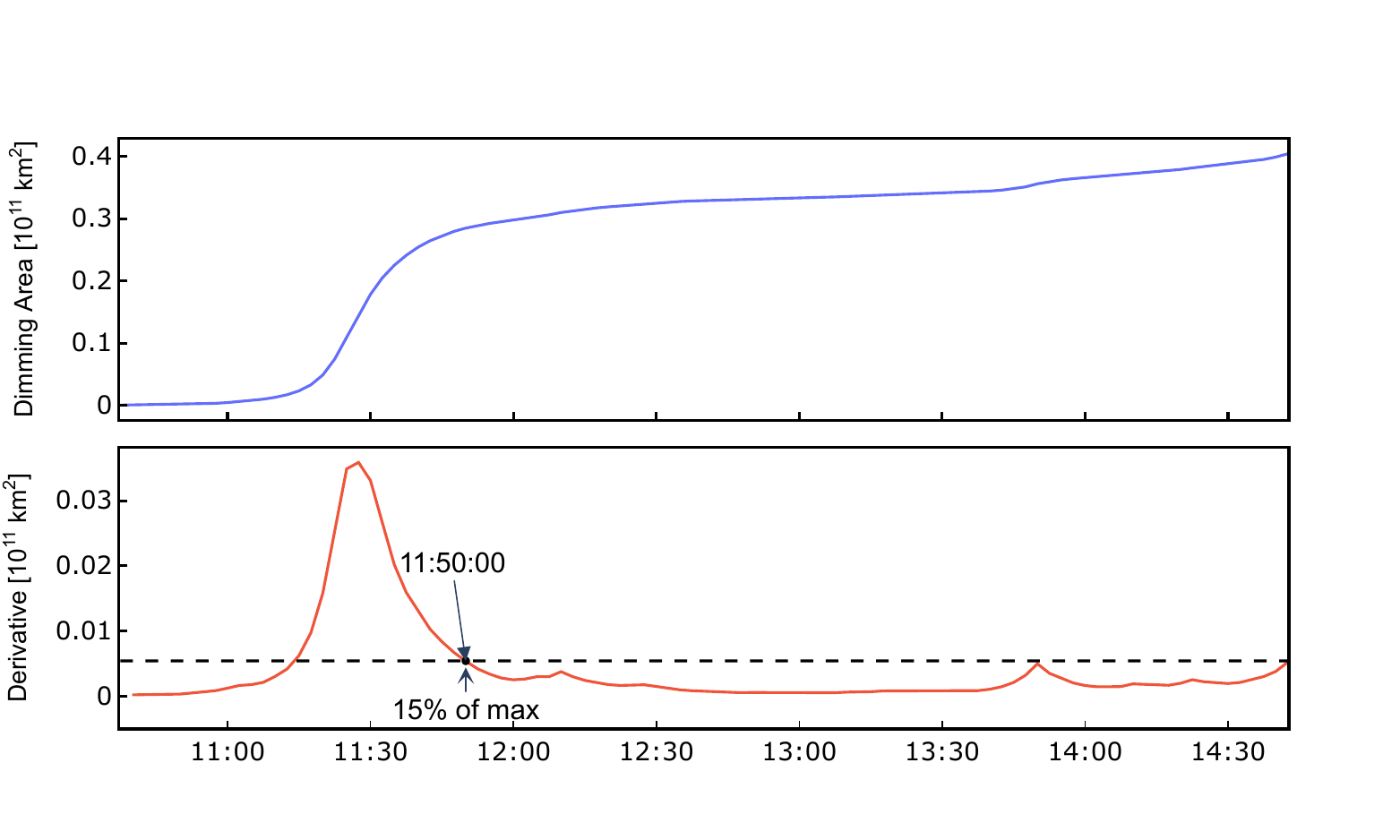}
        \caption*{}
    \end{subfigure}
    
    \begin{subfigure}[b]{0.352\textwidth}
        \centering
        \includegraphics[width=\textwidth]{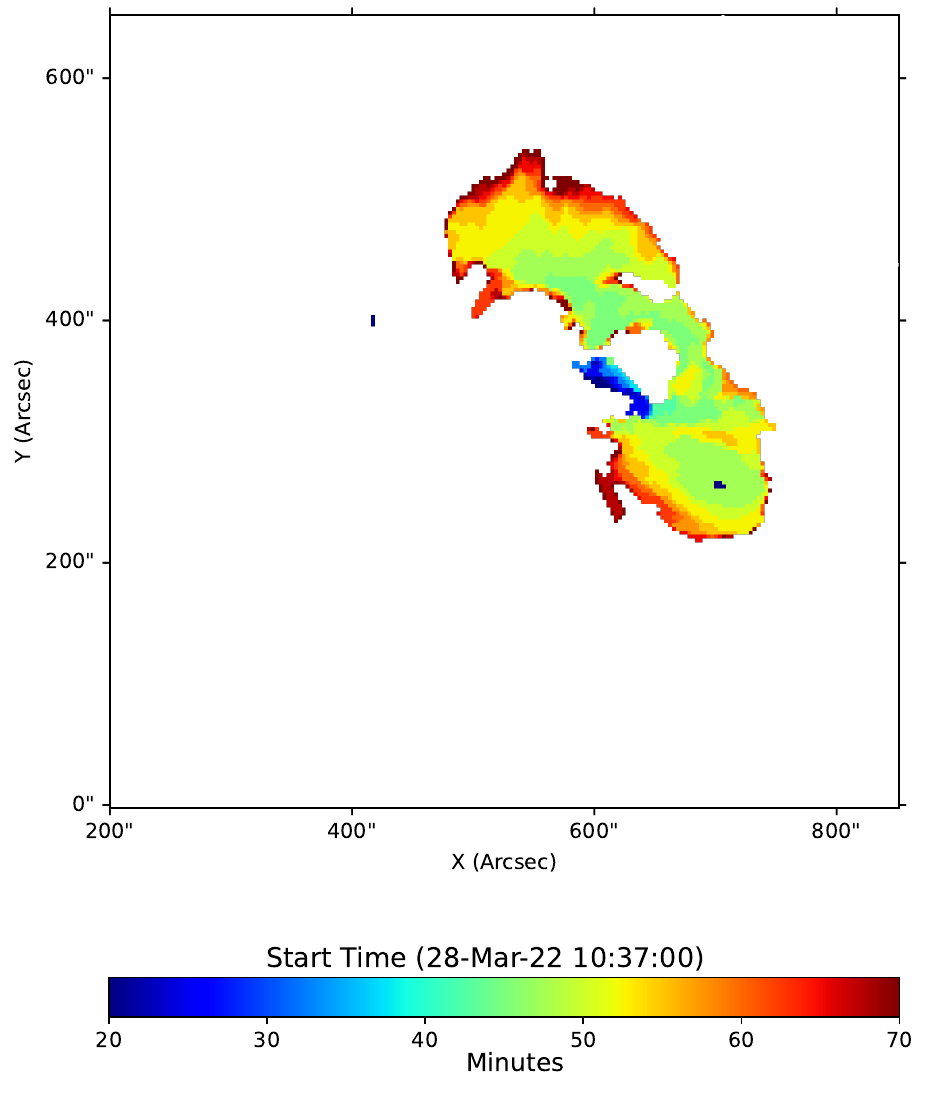}
        \caption*{}
    \end{subfigure}
    
    \begin{subfigure}[b]{0.352\textwidth}
        \centering
        \includegraphics[width=\textwidth]{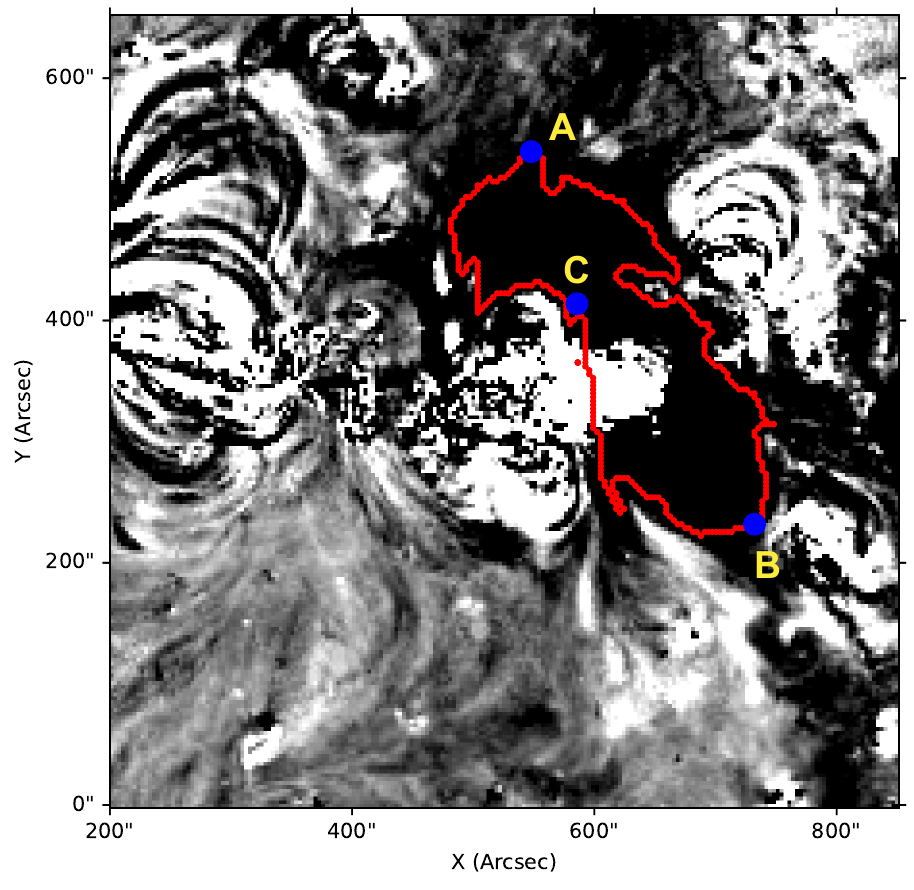}
        \caption*{}
    \end{subfigure}
    \caption{Dimming detection: First row: Expansion of the dimming area $A(t)$ (top panel) and its time derivative $dA/dt$ (bottom panel) over 4 hours. Second row: Cumulative timing map of the dimming extraction, colour-coded in minutes from 10:37 UT using STEREO-A 195~{\AA} logarithmic base ratio images. Third row: STEREO-A 195~{\AA} base-difference image together with the boundary of the identified dimming region (red) at the end of impulsive phase 11:50:00~UT. Point C is the filament source, points A and B mark the largest North and South dimming extent.}
    \label{fig6}
\end{figure}
\begin{figure*}
   \centering
    \begin{subfigure}[b]{0.5\textwidth}
       \centering
        \includegraphics[width=\textwidth]{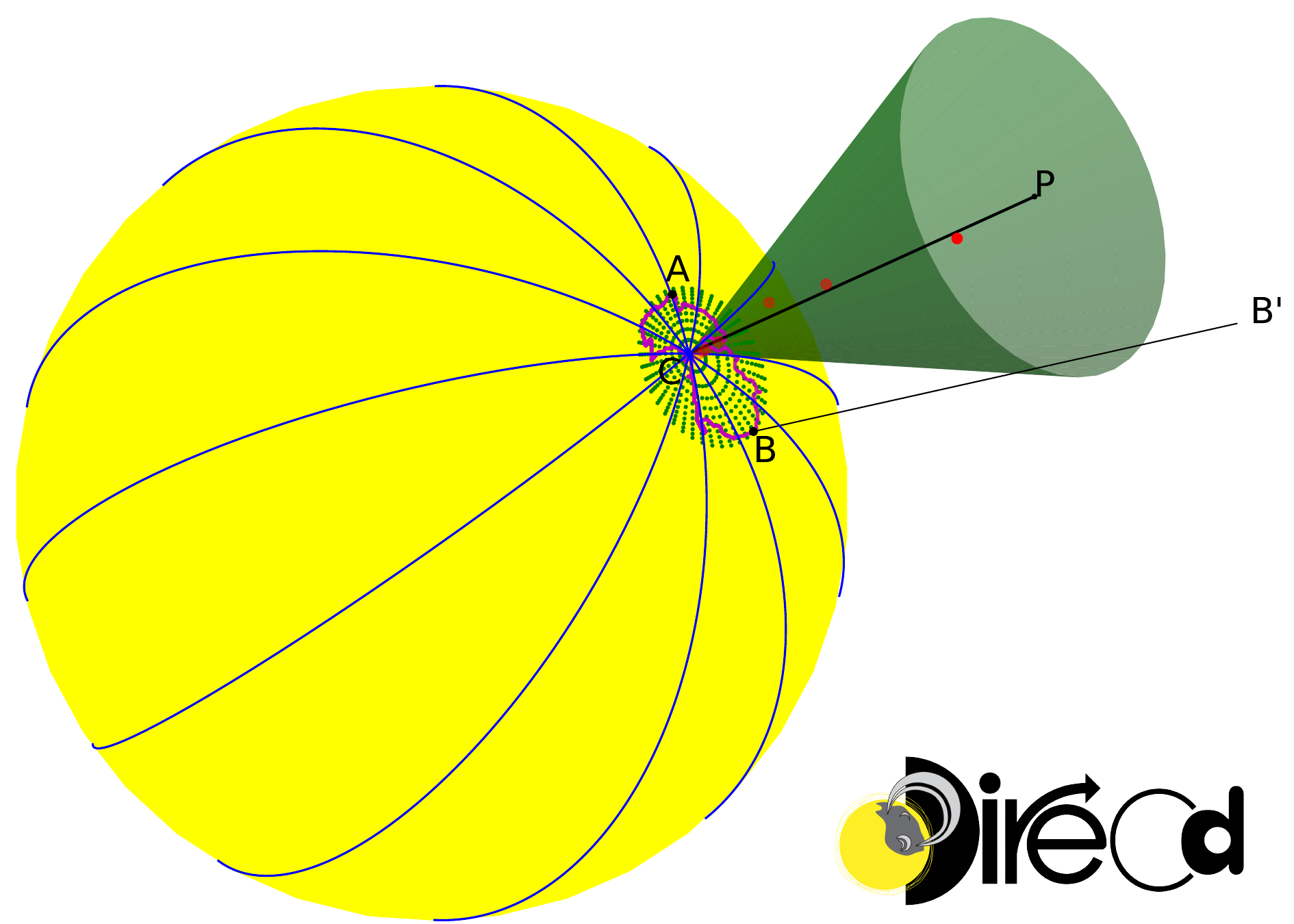}
        \caption*{}
    \end{subfigure}  
    \hspace{0.05\textwidth} 
    \begin{subfigure}[b]{0.33\textwidth}
        \centering
        \includegraphics[width=\textwidth]{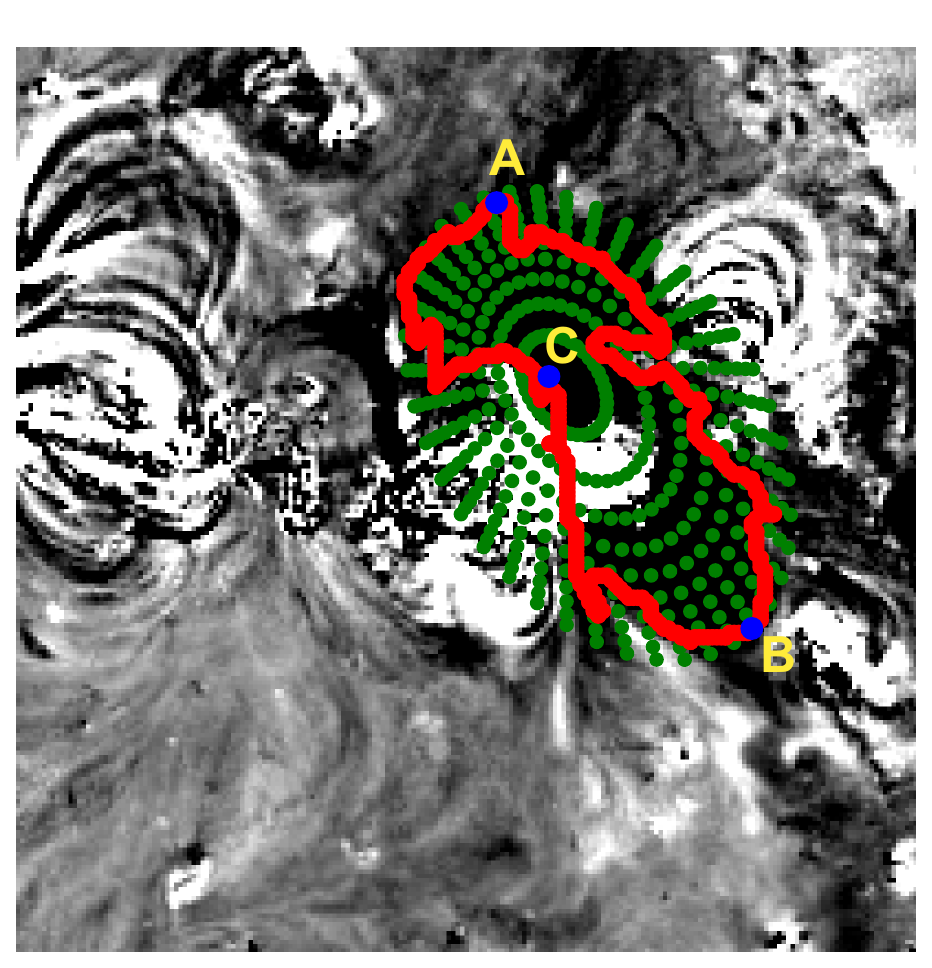}
        \caption*{}
    \end{subfigure} 
    \caption{Application of the DIRECD method to relate the expansion of coronal dimming to the early CME propagation. Left panel: Magenta boundaries outline the dimming.  Point C is filament source, points A and B mark the largest North and South dimming extent. Red markers show the 3D coordinates of the filament, and the black line represents a linear fit to the filament points. The green cone, with a filament fit used as the cone's central axis, has a height of 1.12~$\rm{R_{sun}}$ and a half-width (half-aperture angle) of $21^\circ$. Green dots indicate the orthogonal projections of the CME cone onto the solar surface. We require an edge of the cone base to be orthogonally projected to point B to match the dimming extent. Right panel: STEREO-A 195~{\AA} base-difference image, together with the boundary of the identified dimming region (red) at the end of impulsive phase (11:50:00~UT), and the orthogonal projections (green) of the best fit cone obtained with DIRECD.}
    \label{fig7}
\end{figure*}
\begin{figure*}
   \centering
    \begin{subfigure}[b]{0.45\textwidth}
       \centering
        \includegraphics[width=\textwidth]{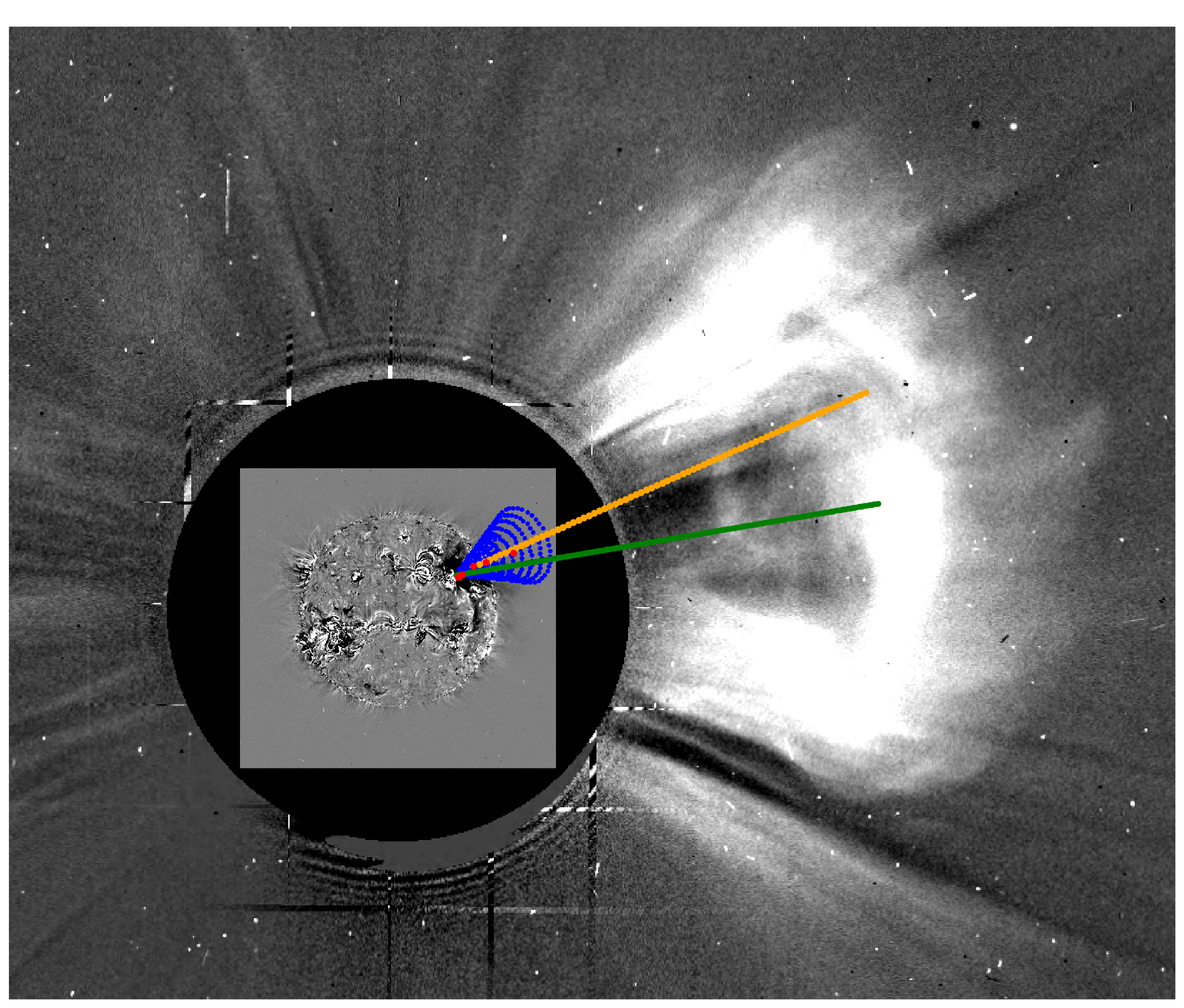}
        \caption*{}
    \end{subfigure}  
    \hspace{0.05\textwidth} 
    \begin{subfigure}[b]{0.427\textwidth}
        \centering
         \raisebox{0.065cm}{ 
            \includegraphics[width=\textwidth]{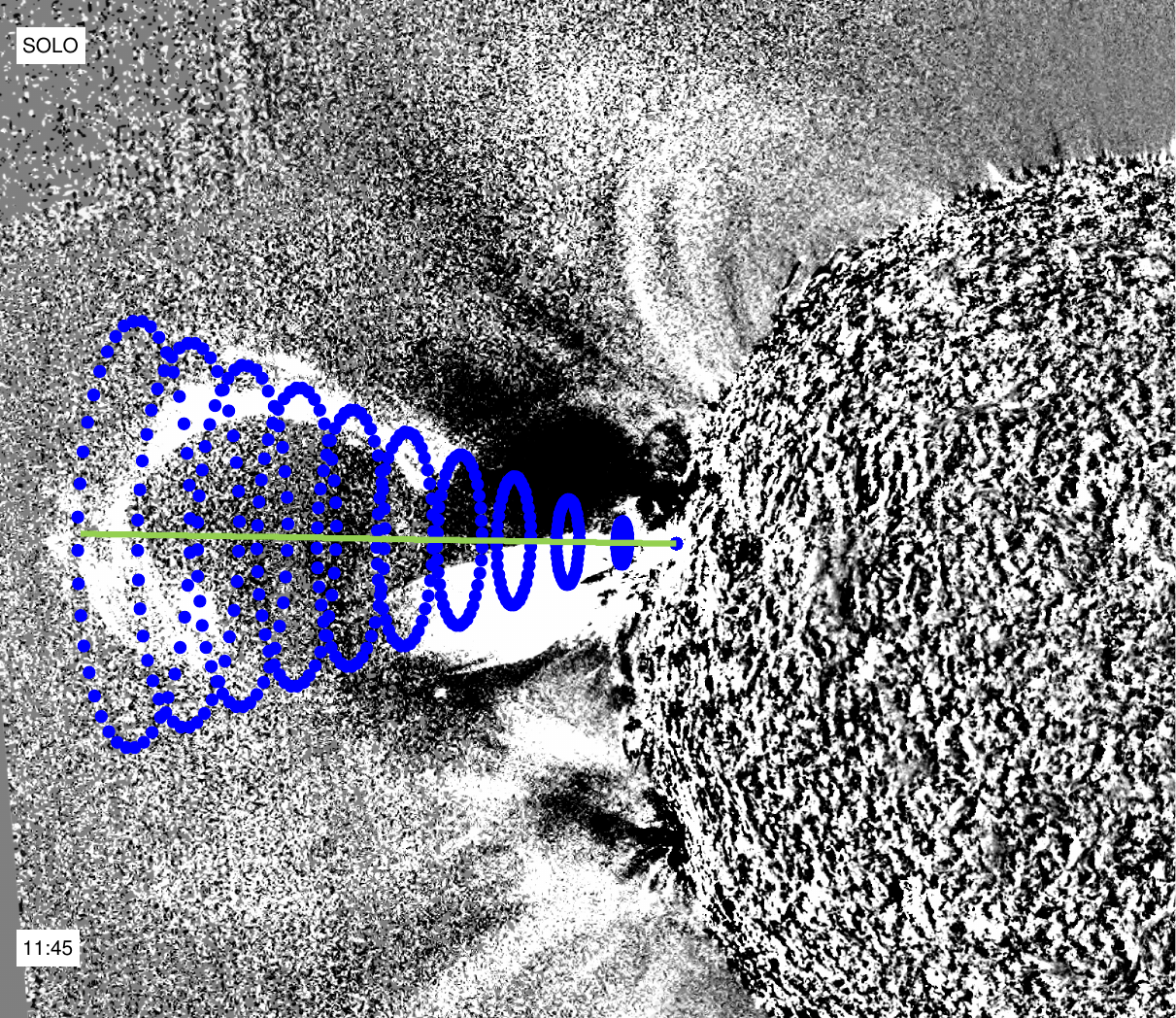}
        }
        \caption*{}
    \end{subfigure} 
    \caption{Filament deflection and 3D cone projections. Left panel: LOS projections of the DIRECD cone (blue mesh) to the STEREO-A view to both 195~\AA~EUVI (11:50~UT) and COR2 (13:23~UT) images. Red dots show the LOS projections of the filament's 3D coordinates, while the orange line represents LOS projections of a linear fit to the filament points used as the cone's central axis (CME pre-deflection direction). The green line indicates the CME post-deflection direction determined from the 3D reconstructions of the filament's height at 11:45~UT (see Figure~\ref{fig4}). Right panel: LOS projections (blue mesh) of the DIRECD cone to 304~\AA~SolO view at 11:45~UT after the filament deflection. The green line shows the CME post-deflection direction. The cone closely matches the filament shape.}
    \label{fig8}
\end{figure*}
\begin{figure*}
   \centering
    \begin{subfigure}[b]{0.45\textwidth}
       \centering
        \includegraphics[width=\textwidth]{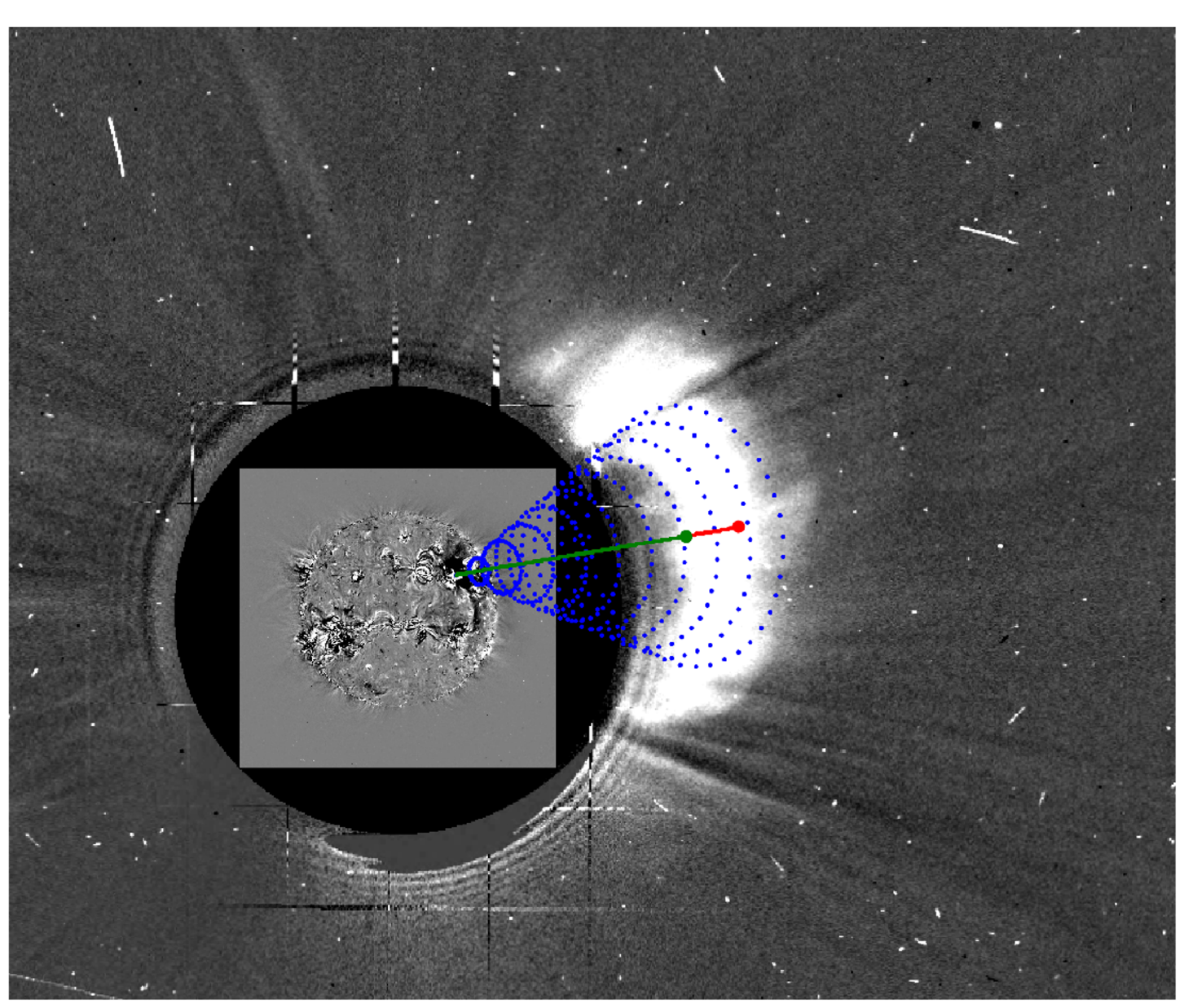}
        \caption*{}
    \end{subfigure}  
    \hspace{0.05\textwidth} 
    \begin{subfigure}[b]{0.45\textwidth}
        \centering
            \includegraphics[width=\textwidth]{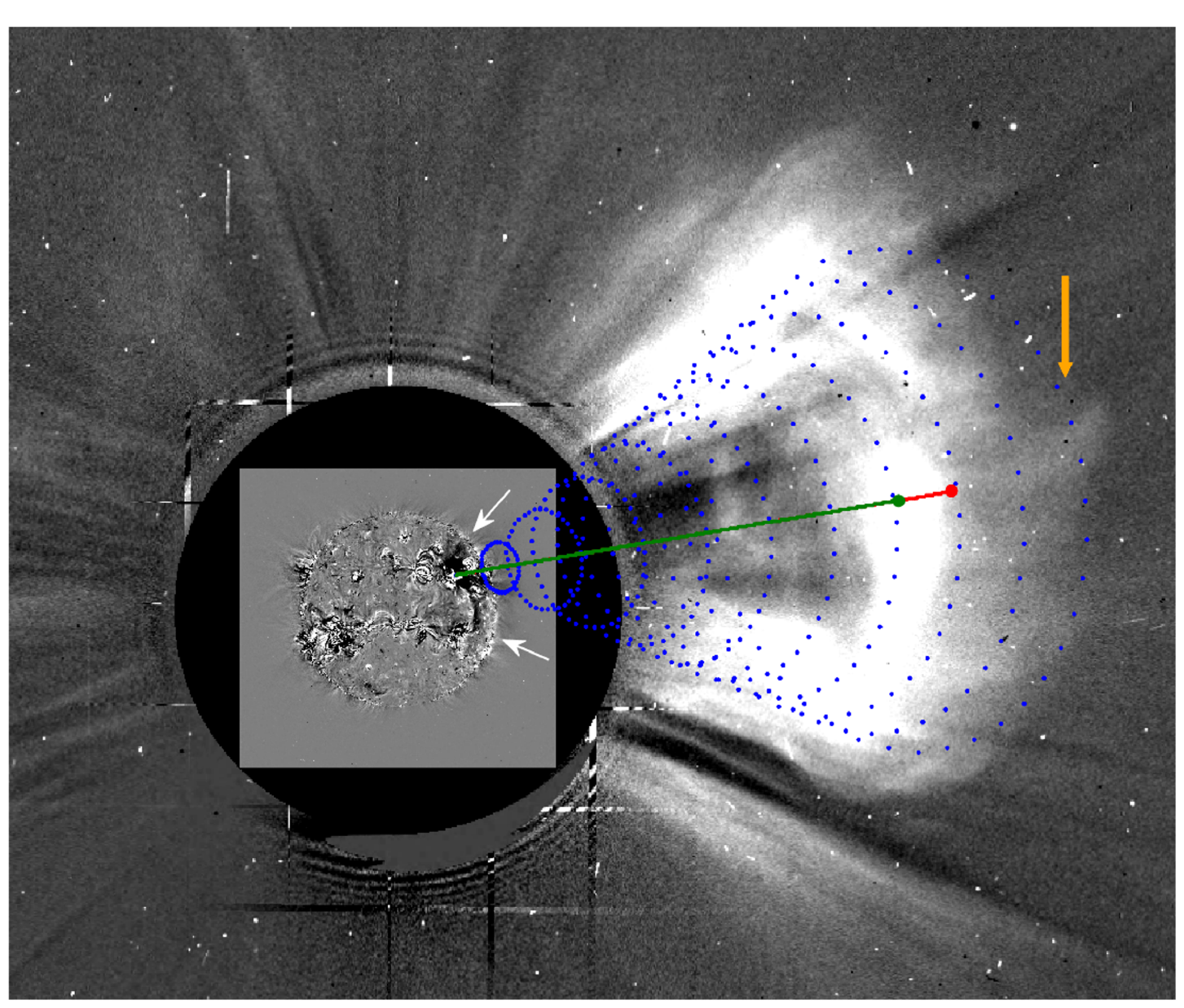}
        \caption*{}
    \end{subfigure} 
    \caption{Extrapolation of the DIRECD cone (blue mesh) to STEREO-A COR2 images at 12:23~UT (left panel) and 13:23~UT (right panel). The green line represents the main axis of the cone, indicating the direction of CME propagation. The green marker indicates the extrapolated height of the filament, while the red marker at the top of the red line indicates the same for the CME front outer edge. The cone approximately  matches the shape of the CME, with fainter CME parts corresponding to far-side cone projections (orange arrow).}
    \label{fig9}
\end{figure*}
Recent studies suggest that coronal dimmings not only indicate the onset of an eruptive event but may also contain information on early CME evolution \citep{Chikunova2023}, leading \citet{Jain2024} to propose the DIRECD method to infer the early CME propagation direction from coronal dimmings.  The DIRECD method uses a 3D cone model to approximate an expanding CME at the end of the dimming's impulsive phase, to estimate key parameters such as direction, half-width, and cone height, for which the CME remains connected to the dimming by matching CME projections on the solar sphere with the observed dimming geometry. In this section, we present the application of the DIRECD method to analyse the early CME evolution by extracting the expansion of the coronal dimming from STEREO/EUVI 195~{\AA} images. Additionally, we validate the resulting CME cone by linking the filament and CME properties derived in the low corona to the white-light coronagraph data.

We detect the coronal dimming in a series of STEREO/EUVI 195~{\AA} images with a 3-minute cadence within the 10:37 -- 14:42~UT time range. The data was calibrated and corrected for differential rotation using the Sunpy library \citep{sunpy_community2020}. We crop the images to the region of interest, examining a subfield around the eruption centre of $x= [-400, 1000]$ and $y= [-300, 1100]$  arcsecs. Dimming pixels are identified in logarithmic base ratio data using the automatic thresholding method described by \cite{Dissauer2018a}.  We extract all pixels with a logarithmic threshold of $-0.19$ and use 30\% of the darkest pixels from this set as seeds for the region-growing algorithm. To minimize the noise and connect the separated parts of the dimming region in STEREO/EUVI images, we also apply median filtering with a $10\times10$ kernel. We save the results of the detection in cumulative dimming maps that include all pixels flagged as dimming up to each time step. To achieve a more accurate estimate of the dimming area on the sphere, especially near the solar limb, we used the method presented in \citep{Chikunova2023}, which allows us to estimate the surface area of a sphere covered by each pixel\footnote{\url{https://github.com/Chigaga/area_calculation}}.

From the evolution of the dimming area, $A(t)$, obtained from these maps (Figure~\ref{fig6}, first row, top panel), we determine the end time of the impulsive dimming growth phase. Following the criteria established by \citet{Dissauer2018b}, this end time is defined as the time when the cumulative dimming area change rate $dA/dt$ has decreased to 15\% of its peak value (Figure~\ref{fig6}, first row, bottom panel). The middle row in Figure~\ref{fig6} illustrates the evolution of the detected cumulative dimming mask during the first 70 minutes following the onset of the event on March 28, 2022, at 10:37~UT. This mask includes all pixels flagged as dimming pixels within this time frame. Each pixel in the cumulative dimming mask is colour-coded based on the time of its initial detection relative to the event's start. Blue regions represent areas identified at the beginning of the event, while red regions indicate areas detected later. The third row in Figure~\ref{fig6} shows a STEREO-A 195~{\AA} base-difference image, along with the boundary of the identified dimming region (in red) at the end of the impulsive phase (11:50:00~UT). The filament source, estimated as the continuation of the filament linear fit to the sphere, is marked by point C. The largest extents of the dimming in North and South are indicated by points A and B, respectively. As shown in the bottom panel of Figure~\ref{fig6}, the dimming is surrounded by several active regions. These include the loop system to the north-west near the dimming boundaries, the active region to the south-west (near point B), and another loop system to the south-east, which partially covers the dimming area.
 
Figure~\ref{fig7} illustrates how we apply the DIRECD method to study the connection between the expansion of the coronal dimming and early CME propagation. To build a 3D CME cone model (green, left panel), we use a linear fit (black line CP) to the reconstructed filament points (red markers) as the central axis of the cone. This axis indicates the direction of CME propagation, with an inclination of $6^\circ$ 
from the radial direction. Additionally, we ensure that the edge of the cone's base is orthogonally projected to point B to match the dimming extent on that side (black line BB\textquotesingle). We then generate an ensemble of 3D cones with varying heights and associated half-widths. From these, we derive the orthogonal projections onto the solar sphere (green dots, left and right panels), which we use as signatures of how the plasma evacuation would be manifested as a coronal dimming when projected onto the solar sphere (for a detailed description of the DIRECD method and its application, we refer to \citep{Jain2024}). 

Figure~\ref{fig1_appendix} displays ten generated CME cones with heights ranging from 0.12 to 1.12~$\rm{R_{sun}}$ and associated half-widths from 20.9 to 65.8$^\circ$ (Cols. 1 and 3), along with their orthogonal projections (blue dots) onto the solar sphere (Cols. 2 and 4) and dimming boundaries (red). The heights in the constructed ensemble of cones represent the values at which the CME, presumed to leave footprints in the low corona up to a certain height, remains connected to the dimming. Figure~\ref{fig2_appendix} presents similar data for CME cones with heights ranging from 1.22 to 2.12~$\rm{R_{sun}}$ and associated half-widths from 15.4 to 19.9$^\circ$. 

In Figure~\ref{fig1_appendix}, it is seen that the orthogonal projections of the CME cones match the dimming extent at point B (as required), extending beyond the dimming boundaries on the side of point A for cones with lower heights. However, as the cone height increases, the projections gradually shrink, moving closer to the dimming boundaries. The first instance where point A is the closest to the projections occurs for the cone with a height of 1.12~$\rm{R_{sun}}$ and a half-width of $21^\circ$. Additionally, point A falls within the CME cone projections for the cone with a height of 1.22~$\rm{R_{sun}}$ and a half-width of $19.9^\circ$ (Figure~\ref{fig2_appendix}). However, as the cone height increases further, the dimming boundary extends beyond the projection. Consequently, we identify only two cones with heights of 1.12 and 1.22~$\rm{R_{sun}}$, and associated half-widths of 20.9 and $19.9^\circ$, respectively. In these cases, point A falls within the projections, and the entire dimming area remains encompassed. We chose the projection of a cone with a height of 1.12~$\rm{R_{sun}}$ and a half-width of $21^\circ$ onto the solar surface as the best match for the dimming geometry, with point A being the closest to the cone projection boundaries (shown in green in the left panel of Figure~\ref{fig7}). As can be seen in the right panel of Figure~\ref{fig7}, the orthogonal projections (green) of the best-fit cone obtained with DIRECD in a STEREO-A {195~\AA} base-difference image closely match the dimming geometry at the end of the impulsive phase (11:50:00~UT), including both the identified dimming region (red) and the dimming areas covered by active regions and loop systems. 

As demonstrated in \citet{Jain2024}, the shape of the CME cone projection varies from circles to ellipses depending on the inclination angle from the radial direction and the source location. For a radial 3D cone, the CME projections form concentric circles around the source near disk centre, whereas they transform into ellipses when the source is located close to the solar limb.  As seen in both panels of Figure~\ref{fig7}, the shape of the 3D cone projection (green mesh) as well as that of the dimming in EUV resembles an ellipse. Since the CME originates from close to the solar limb and propagates nearly radially, we suggest that the dimming shape and source location could also serve as preliminary indicators of the direction of early CME propagation.
  
To link the structures of the eruption that we observe in the EUV with the traditional white-light coronagraph data, we show in the left panel of Figure~\ref{fig8} LOS projections of the DIRECD cone from the STEREO-A view for both the 195~\AA~EUVI (11:50~UT) and the coronagraph data (13:23~UT), capturing the moment when the CME is visible in COR2. Red dots indicate LOS projections of the filament 3D coordinates, whereas the orange line represents LOS projections of a linear fit to the filament points, used as the cone's central axis (CME pre-deflection direction). While the central axis of the cone exhibits a slight offset from the observed centre of the CME bubble (the outer CME envelope) in COR2, the green line representing the CME post-deflection direction (see Figure~\ref{fig4}) directly targets the bubble's centre.

We may speculate that the deflection was caused by the loop system to the north-west near the dimming boundaries (see the right panel of Figure~\ref{fig7}). In a recent study, \citet{Chikunova2023} showed that the dominant propagation of dimming growth may indicate the propagation direction of the erupting filament. However, since the deflection in this case was observed towards the end of the dimming's impulsive phase (see the first row of Figure~\ref{fig6}), when most of the dimming region had already developed, minor changes in the dimming evolution are not prominent enough to trace changes in the filament propagation direction. However, we assume that the connection between the dimming and the expanding CME is fully established by the end of the dimming impulsive phase. Thus, we further construct a new CME cone using the filament's post-deflection direction as the cone's central axis. This new cone maintains the same height (1.12~$\rm{R_{sun}}$) and has a half-width of $21^\circ$, derived from the dimming before the deflection. As it can be seen in the right panel of Figure~\ref{fig8}, the reconstructed cone (blue mesh) closely aligns in height and angular width with the observed filament shape, which represents the inner part of the CME.

To ensure that the filament and CME properties derived in the low corona correspond to the white-light structure observed in COR2, we extrapolate the filament height kinematics (Figure~\ref{fig3}) derived from the SolO, STEREO-A and SDO EUV data to the COR2 FOV. By applying a linear fit to the evolution of the filament in Figure~\ref{fig3}a (blue line) for data after 11:19:29~UT, we determine a filament height of 3.60~$\rm{R_{sun}}$ at 12:23~UT (green line with green marker on top in Figure~\ref{fig9}, top panel) and 6.94~$\rm{R_{sun}}$ at 13:23~UT (green line with green marker on top in Figure~\ref{fig9}, bottom panel). Considering that the distance between the filament and the CME front outer edge along the filament propagation direction more than doubled in the low corona over 10 minutes (from around 93 to 212~Mm, as noted in Figure~\ref{fig5}, right panel, green line, from 11:25 to 11:35 UT), we also extrapolate the CME front outer edge to the COR2 FOV. We obtain a distance of 4.43~$\rm{R_{sun}}$ at 12:23~UT (red line with red marker on top in Figure~\ref{fig9}, left panel) and 7.76~~$\rm{R_{sun}}$ at 13:23~UT (red line with red marker on top in Figure~\ref{fig9}, right panel).  
 
We further construct expanded CME cones (blue mesh in Figure~\ref{fig9}) using the extended filament as the cone's central axis. These cones maintain the same half-width of $21^\circ$ (derived from the dimming before the deflection) and have heights of 3.60~$\rm{R_{sun}}$ at 12:23~UT (left panel) and 6.94~$\rm{R_{sun}}$ at 13:23 UT (right panel). As seen in Figure~\ref{fig9}, the extrapolated filament heights align closely with the inner parts of the CME bubble, while the extrapolated CME front outer edge matches the edge of the CME structure observed at two different time steps in COR2. Additionally, the extended cones closely match the shape of the CME, with the fainter CME parts closely corresponding to far-side cone projections (indicated by the orange arrow in the bottom panel). This correspondence furthermore indicates that analysing the dimming evolution during its impulsive phase, reconstructing the filament in 3D, and estimating its kinematics allowed us to extract key CME parameters early in its development—such as angular half-width and propagation direction—revealing a self-similar evolution following the filament's deflection at 11:45~UT. We note that remote, secondary dimming regions further extend to the North and South of the initial, central dimming region. These are not fitted within DIRECD (see white arrows in Figure~\ref{fig9}, right panel) but correspond to the outer edges of the CME bubble's white-light signature as observed in the STEREO-A COR2 coronagraph.

\section{Discussion and conclusions}
The study of the eruptive event on March 28, 2022, reveals a three-part structure of the CME rarely observed in the low corona, including a bright core, dark cavity, and bright front edge. Our analysis using data from three well-separated viewpoints (SolO, STEREO-A, SDO) shows that the outer front observed in SolO/EUI 304 {\AA} filtergrams (with the erupting filament embedded inside) corresponds to the shock structures in STEREO-A 195 {\AA}, where we observe the full 3D EUV wave-dome suggestive of the coronal shock \citep{veronig2010}, and to the outer CME front that appears later in STEREO-A COR2. The shock scenario is further supported by the associated radio type II burst, which fits in time and location to the EUV wave dome \citep{Morosan2024}.

Three spacecraft, SolO, STEREO-A, and SDO, observed the filament eruption from different perspectives, allowing us to conduct a detailed analysis of its path, height, and kinematics. The filament's height increased dramatically from 28 to 616~Mm (0.04 to 0.89~$\rm{R_{sun}}$) in just 30 minutes (11:05-11:35~UT). Its speed reached  $648 \ \pm 51 \ \rm{km} \ \rm{s}^{-1}$,  a peak acceleration of $1624 \ \pm 332 \ \rm{m} \ \rm{s}^{-2}$. This eruption was associated with an M4-class solar flare, with the impulsive filament acceleration starting and peaking several minutes before the first (main) HXR burst observed by STIX. This timing suggests that, in this event, the eruption triggered the flare, leading to magnetic reconnection and particle acceleration.

We developed the ATLAS-3D method, validated against traditional 3D approaches, to derive the filament's 3D coordinates and height at later stages using only data from the SolO spacecraft, integrated with earlier 3D reconstructions. This method revealed a filament deflection around 11:45~UT, reaching a height of 841~Mm (1.21~$\rm{R_{sun}}$), with 3D reconstructions of filament loops and the CME shock structure establishing their spatial relationship. The bright CME leading edge grew from 383 to 837~Mm (0.55 to 1.2~~$\rm{R_{sun}}$) between 11:25 and 11:35~UT, with the distance between the filament apex and CME leading edge more than doubling from 93 to 212~Mm (0.13 to 0.3~$\rm{R_{sun}}$), indicating substantial expansion and increase of the spatial separation. These findings may indicate that after the bow shock formation, it continues propagating freely, whereas the filament expansion slows down.

Using the DIRECD method to explore the relationship between coronal dimming, filament eruption, and early CME propagation, we found that connections between dimming and CME expansion are fully established by the end of the dimming's impulsive phase, preceding the filament's deflection at 11:45~UT, and revealing subsequent self-similar CME evolution. Key parameters estimated from DIRECD and 3D filament reconstructions include the CME direction aligned with the filament motion (inclined by $6^\circ$ from radial direction, a half-width of $21^\circ$, and a cone height of $1.12~R_{\text{sun}}$, for which the CME shows connections to the dimming and still leaves footprints in the low corona. Notably, the reconstructed 3D CME cone closely aligns in height and angular width with the observed filament shape at 11:45~UT, representing the inner part of the CME. This finding aligns with \citet{Chikunova2023, Jain2024}, who demonstrated that the dimming morphology closely mirrors the inner structure of the 3D GCS CME reconstructions (the bottom part of the GCS croissant), supporting the hypothesis that CME propagation is linked to dimming extent and leaves distinct imprints in the low corona up to a limited height, as shown by the DIRECD-derived parameters.

We validated the 3D CME cone by correlating CME and filament properties from the low corona with white-light coronagraph data. Extrapolating filaments and CME heights to STEREO-A COR2 times showed filament heights closely align with the inner CME bubble, while the CME front matches COR2 observations. Extended 3D cones also resemble the CME shape, with fainter parts aligning with far-side projections. Linking expanding dimming to early CME stages, this approach underscores coronal dimmings as valuable indicators for early tracking of CME evolution and demonstrates that the DIRECD method not only correlates 2D dimming with the 3D CME bubble but also supplements GCS reconstructions with additional information in the lower corona.

In summary, we demonstrate that the bright core of the CME three-part structure in this event corresponds to the filament/prominence, whereas the leading front corresponds to the shock. Furthermore, the (extended) filament/prominence corresponds to what we later observe as a CME in white light coronagraphs. The findings and outcomes of this study offer valuable insights into early CME evolution, showcasing the importance of multi-viewpoint observations and novel reconstruction methods for improving space weather forecasting and mitigating its impacts.

\begin{acknowledgements}
T.P., S.J. acknowledge support by the Russian Science Foundation under the project 23-22-00242, \url{https://rscf.ru/en/project/23-22-00242/}. M.D. and G.C. acknowledge support by the Croatian Science Foundation under the project IP-2020-02-9893 (ICOHOSS) and support from the Austrian-Croatian Bilateral Scientific Project ``Analysis of solar eruptive phenomena from cradle to grave”. A.M.V. and S.P. acknowledge the Austrian Science Fund (FWF) 10.55776/I4555. Solar Orbiter is a space mission of international collaboration between ESA and NASA, operated by ESA. The STIX instrument is an international collaboration between Switzerland, Poland, France, Czech Republic,Germany, Austria, Ireland, and Italy. The EUI instrument was built by CSL, IAS, MPS, MSSL/UCL, PMOD/WRC, ROB, LCF/IO with funding from the Belgian Federal Science Policy Office (BELSPO/PRODEX PEA 4000134088); the Centre National d’Etudes Spatiales (CNES); the UK Space Agency (UKSA); the Bundesministerium für Wirtschaft und Energie (BMWi) through the Deutsches Zentrum für Luft- und Raumfahrt (DLR); and the Swiss Space Office (SSO). SDO data are courtesy of the NASA/SDO AIA and HMI science teams. The STEREO/SECCHI data are produced by an international consortium of the Naval Research Laboratory (USA), Lockheed Martin Solar and Astrophysics Lab (USA), NASA Goddard Space Flight Center (USA), Rutherford Appleton Laboratory (UK), University of Birmingham (UK), Max-Planck-Institut für Sonnenforschung (Germany), Centre Spatiale de Liège (Belgium), Institut d’Optique Théorique et Appliquée (France), and Institute d’Astrophysique Spatiale (France). 
We thank the referee for valuable comments on this study.
\end{acknowledgements}

\bibliographystyle{aa}
\bibliography{My_References}

\begin{appendix}
\section{Ensemble of cones}\label{Appendix_A}

\noindent
\begin{minipage}{\textwidth}
    \centering
    \includegraphics[width=15cm]{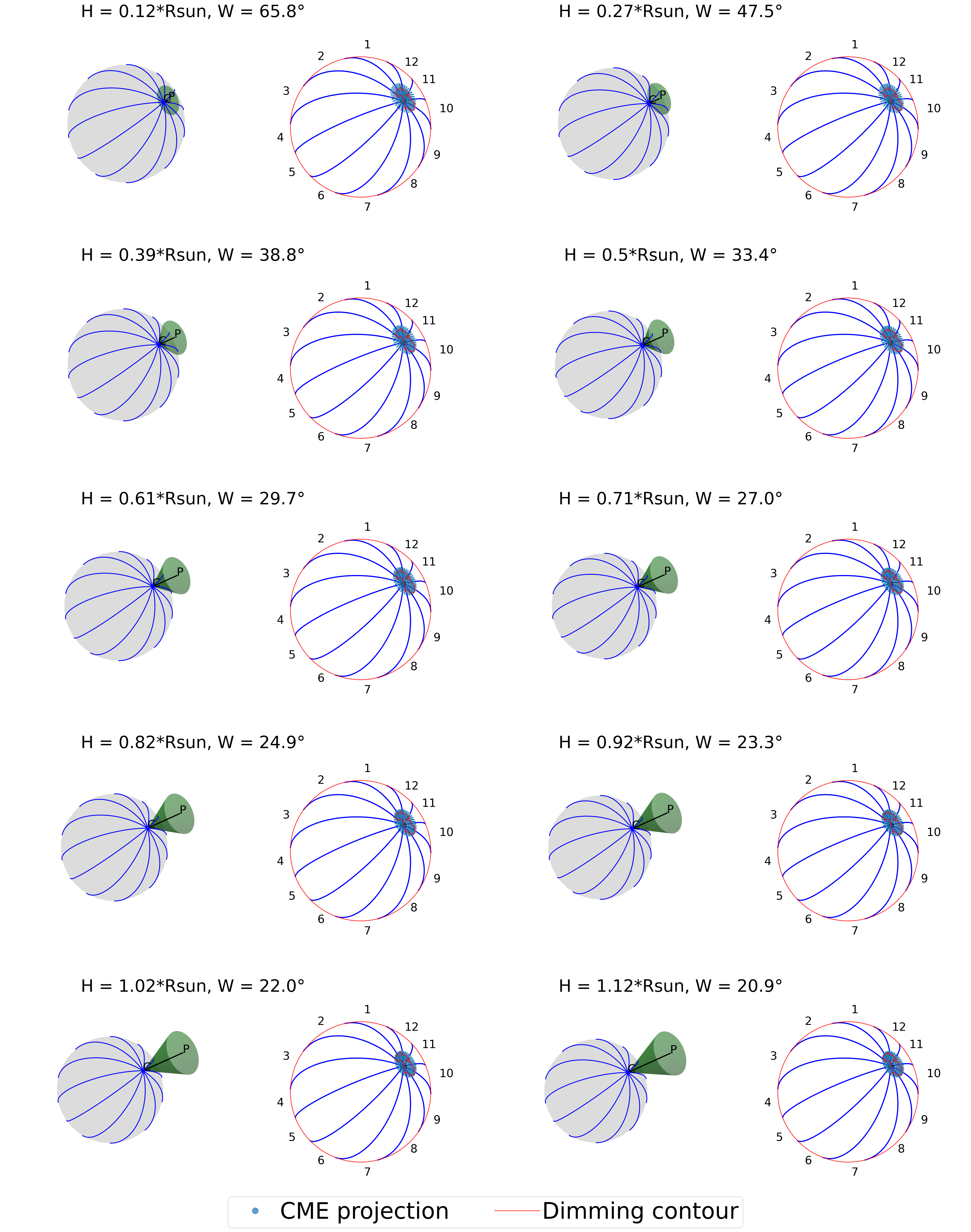}
    \captionof{figure}{CME cones at heights of 0.12–1.12~$\rm{R_{sun}}$ with associated half-widths of 20.9-65.8$^\circ$ (Cols. 1 and 3) and their orthogonal projections (blue dots) onto the solar sphere (Cols. 2 and 4). One edge of the cone is bounded by the OB line, where point O is the Sun's centre and point B is the dimming edge (see Figure~\ref{fig7}). The dimming boundaries are outlined in red. The inclination angle for all the cones is 6$^\circ$. Columns 1 and 3 depict the side view to better show the reconstructed cones, and Cols. 2 and 4 show the face-on view.}
    \label{fig1_appendix}
\end{minipage}

\begin{figure*}[h]
	\centering
	\includegraphics[width=15cm]{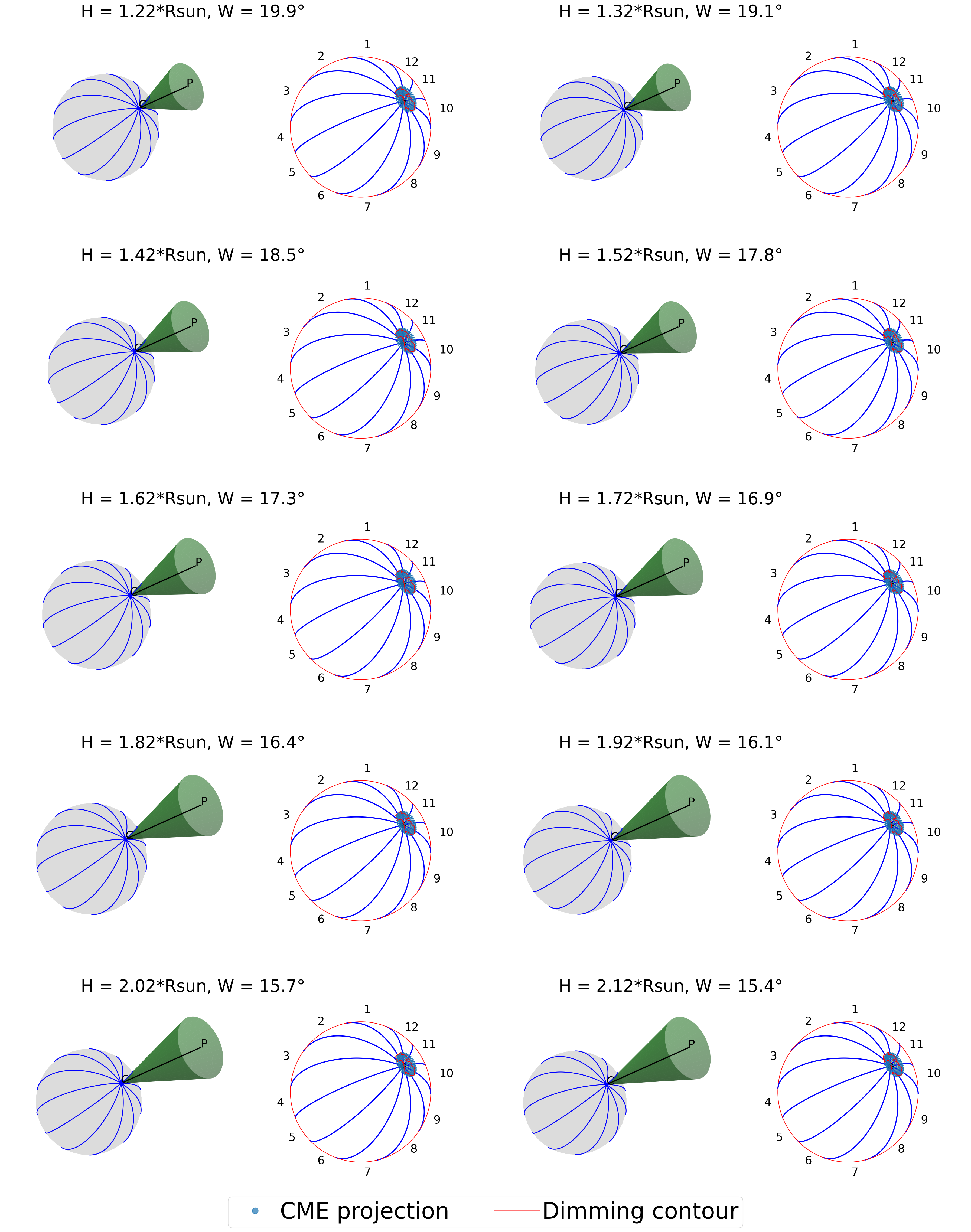} 
	\caption{The same as Figure~\ref{fig1_appendix}, but for CME cones at heights of 1.22--2.12~$\rm{R_{sun}}$ with associated half-widths of 15.4--19.9$^\circ$.}
	\label{fig2_appendix}
\end{figure*}

\end{appendix}
\end{document}